\documentclass{aa}
\usepackage{natbib}
\usepackage{graphicx}
\usepackage{txfonts}
\usepackage[dvipsnames]{xcolor}
\usepackage{amsmath}
\usepackage{comment}
\usepackage[colorlinks=true, linkcolor=blue, citecolor=blue, urlcolor=blue]{hyperref}
\newcommand{\btheta}{{\boldsymbol{\theta}}}

\begin{document}

   \title{Learning cosmic web environments with diffusion models}

   \author{M.~Noor\inst{1, 2}\thanks{\footnotesize Corresponding author: \href{mailto:mehdi.noor@universite-paris-saclay.fr}{\scriptsize \nolinkurl{mehdi.noor@universite-paris-saclay.fr}}}
        \and T.~Bonnaire\inst{1, 3}
        \and N.~Aghanim\inst{1}
        \and A.~Decelle\inst{4, 5}
}

\institute{Université Paris-Saclay, CNRS, Institut d'Astrophysique Spatiale, 91405 Orsay, France \and Centre national d’études spatiales (CNES), France
           \and Laboratoire de Physique de l’École normale supérieure, ENS, Université PSL, CNRS, Sorbonne Université, Université Paris Cité, F-75005 Paris, France
           \and Escuela Técnica Superior de Ingenieros Industriales, Universidad Politécnica de Madrid, Calle de José Gutiérrez Abascal 2, Madrid 28006, Spain
           \and GISC - Grupo Interdisciplinar de Sistemas Complejos 28040 Madrid, Spain
}
 
  \abstract{
  The cosmic web, consisting of an intricate network of voids, walls, filaments, and nodes, encodes key information about structure formation and the cosmological parameters that govern it. In the era of high-precision cosmology, large ensembles of numerical simulations are required to analyse next-generation galaxy surveys, motivating the use of generative models to circumvent their high computational cost. While they have shown promise in emulating high-fidelity cosmic web simulations, their ability to capture distinct cosmic web environments remains largely unexplored. For this study, we trained a diffusion model on the \textsc{Quijote} $N$-body simulation suite to investigate the semantic information learnt by its self-attention maps. Using statistical estimators such as the Dice coefficient and cross-power spectra, we quantified the correspondence between attention maps and cosmic web environments defined by the T-Web classifier. We find that attention maps of varying spatial resolutions across different layers capture overdense and underdense structures in distinct ways, exhibiting strong positive correlations and anti-correlations with both the overall matter distribution and individual cosmic web environments. Moreover, the diffusion model predominantly encodes cosmological information at intermediate-to-large spatial scales, indicating that attention maps primarily capture globally coherent structures. 
  Our results show that, beyond accurately reproducing two-point statistics, diffusion models learn a multi-scale representation of the cosmic web through self-attention, including non-Gaussian information.}


   \maketitle

\section{Introduction}
   In the early Universe, primordial fluctuations in the energy density grew by undergoing anisotropic gravitational collapse, which led to an inhomogeneous distribution of dark matter (DM) and galaxies that defines the large-scale structure of the Universe today. On megaparsec scales, this distribution traces out a three-dimensional, filamentary pattern, forming an interconnected network known as the cosmic web \citep{Bond_1996}, composed of four main environments. Voids are vast underdense regions that expand to dominate the volume of the cosmic web. Their boundaries are defined by sheet-like structures of matter called walls. From these walls, matter flows into elongated structures known as filaments, which emerge at the intersection of walls. Matter then flows along these filaments and accretes at their intersection to form dense nodes. These spherical overdensities host numerous DM halos and massive galaxy clusters \citep[e.g. ][]{Cautun_2014}. 
   
   The cosmic web pattern has been consistently revealed by large galaxy surveys, which map the spatial distribution of galaxies in the Universe \citep[e.g. ][]{1989Sci...246..897G,colless20032dfgalaxyredshiftsurvey,Tegmark_2004,2005ASPC..329..135H}. Uncovering the cosmological information encoded in the cosmic web is essential. However, the precision of cosmological parameter constraints from galaxy surveys is limited by our ability to model non-linear scales and the baryonic effects that impact galaxies as biased tracers of the underlying DM distribution. State-of-the-art galaxy surveys, such as \textsc{Euclid}\footnote{\url{https://www.esa.int/Science_Exploration/Space_Science/Euclid}}, will provide unprecedented high-precision measurements of observables, increasing the demand for accurate modelling of the non-linear regime. Early numerical efforts to model the large-scale matter distribution, motivated by the Zel'dovich approximation 
   \citep{1970A&A.....5...84Z}, include the seminal works of, for example, \citet{1980MNRAS.192..321D}, \citet{1983MNRAS.204..891K}, and \citet{Jenkins_1998}, which reproduced the filamentary cosmic web pattern. More recent studies have simulated the formation and evolution of the cosmic web from primordial density fluctuations at substantially higher mass resolutions, including hydrodynamical simulations that account for complex baryonic physics \citep[e.g. ][]{Schaye_2023, Hern_ndez_Aguayo_2023}.

    Despite the associated computational cost, numerical simulations are necessary to interpret observational data or construct accurate covariance matrices for likelihood analyses. Additionally, approaches such as simulation-based inference \citep{Cranmer_2020} further increase this demand, requiring large ensembles of simulations to train neural networks effectively, in the order of at least $\sim 10^{3}-10^{4}$ simulations \citep{Bairagi_2025}. Furthermore, the accurate modelling of baryonic effects at non-linear scales where linear perturbation theory breaks down is, in itself, computationally expensive, requiring tens of thousands of CPU hours \citep{Schneider_2016}. Over the past decade, these challenges have motivated the use of generative models to emulate high-fidelity simulations in order to explore the possibility of mitigating this computational bottleneck. Large suites of $N$-body simulations, particularly \textsc{Quijote} \citep{Villaescusa_Navarro_2020}, have been developed to quantify the information content of cosmological observables, while also providing comprehensive datasets for training machine-learning algorithms.

   In recent years, generative models have made rapid advancements in realistic data generation, including restricted Boltzmann machines \citep[RBMs,][]{hinton2002training,carbone2024fast,bereux2024fast}, variational autoencoders \citep[VAEs,][]{rezende2014stochasticbackpropagationapproximateinference, kingma2022autoencodingvariationalbayes}, autoregressive models \citep{germain2015mademaskedautoencoderdistribution, oord2016pixelrecurrentneuralnetworks}, and generative adversarial networks \citep[GANs,][]{goodfellow2014generativeadversarialnetworks}, which were among the first widely adopted generative models for emulating cosmological data. They have shown promising results in cosmology for emulating $N$-body simulations \citep{Rodr_guez_2018, Feder_2020, Ullmo_2021}. However, GANs have several limitations, including training instabilities and the well-known issue of mode collapse, where the generator of the model produces samples from only a few modes of the underlying data distribution, resulting in highly similar or nearly identical samples \citep{salimans2016improvedtechniquestraininggans, thanhtung2020catastrophicforgettingmodecollapse, Ullmo_2021}.

   More recently, diffusion probabilistic models \citep[hereafter diffusion models,][]{sohldickstein2015deepunsupervisedlearningusing, ho2020denoisingdiffusionprobabilisticmodels} have emerged as a powerful alternative for capturing complex data distributions and producing realistic samples. They approximate the underlying data distribution by learning its score function, that is, the gradient of the logarithm of the data probability density. This is achieved by means of a denoising process that progressively transforms random noise into high-fidelity samples with resolved structures. Diffusion models also demonstrate greater training stability, avoid mode collapse and achieve diverse, high-fidelity samples, while effectively generalising beyond their training data \citep{dhariwal2021diffusionmodelsbeatgans, bonnaire2025diffusionmodelsdontmemorize}.
   
   In cosmology, diffusion models have already been applied across a variety of tasks, from generating realistic mock galaxy images \citep{Smith_2022} and conditional $21~\mathrm{cm}$ brightness temperature maps \citep{zhao2023diffusionmodelconditionallygenerate}, to the emulation of $N$-body simulations conditioned on cosmological parameters \citep{mudur2023cosmologicalfieldemulationparameter}, and also enhancing simulations through super-resolution techniques \citep{Schanz_2024, rouhiainen2024superresolutionemulationlargecosmological, mishra2026cosmofoldfastgenerationupscaling}. With increasing use of generative models to emulate cosmological data, we present a study that explores how diffusion models learn to capture distinct cosmic web environments. While previous studies have demonstrated that diffusion models can reproduce the statistical properties of cosmological simulations, their internal mechanisms for capturing cosmic web structures have not been explored in depth. In particular, the role of self-attention maps \citep{vaswani2023attentionneed}, which are learnt neural network features embedded within the architecture that capture important multi-scale semantic information, remains largely unexplored in this context.
   
   To understand how diffusion models learn the cosmic web, we trained a diffusion model, which incorporates self-attention layers throughout its architecture, on two-dimensional patches of log-transformed DM density fields from the \textsc{Quijote} suite. We performed both a qualitative and quantitative analysis of its self-attention maps of different spatial resolutions to evaluate how sensitive they are to both the overall matter structure in the cosmic web and the individual cosmic web environments. For this, we adopted a dynamical classifier that classifies the four main cosmic web environments. These segmented fields act as our ground truth to assess the cosmological information captured by these attention maps. This analysis provides insight into how diffusion models capture important structural information from cosmic web environments across multiple scales. The ability of attention maps to capture such multi-scale, long-range dependencies is essential for reconstructing the non-Gaussian structures of the cosmic web and its higher-order statistics, which are increasingly required for precise cosmological analyses.

   This paper is organised as follows. Sect.~{\ref{sec_methodology}} briefly introduces the theory of diffusion models and their application to training a model that accurately reproduces the statistics of DM simulations. We also discuss self-attention maps and the algorithm used to produce segmentation masks of DM density fields. This is then followed by a description of the statistical estimators used in our analysis in Sect.~\ref{sec_stat_estimators}, followed by the results obtained from both qualitative and quantitative analyses of attention maps in Sect.~{\ref{sec_results}}. We conclude with a discussion and summary of our work in Sect.~{\ref{sec_discussion}} and Sect.~{\ref{sec_conclusion}}.

\section{Methodology}\label{sec_methodology}

    \begin{figure*}[htbp]
        \centering
        \includegraphics[width=0.734\textwidth]{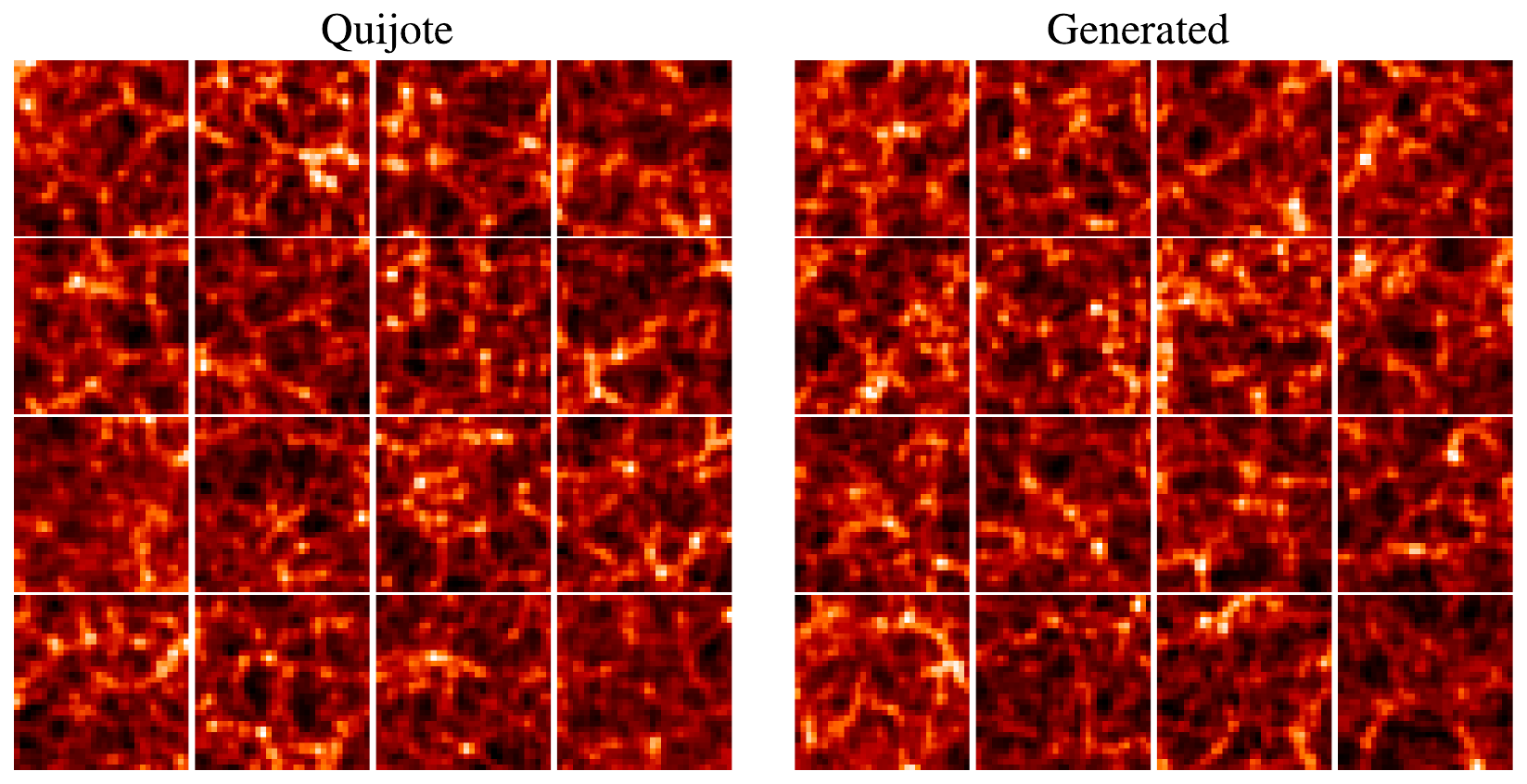}
        \includegraphics[width=0.26\textwidth]{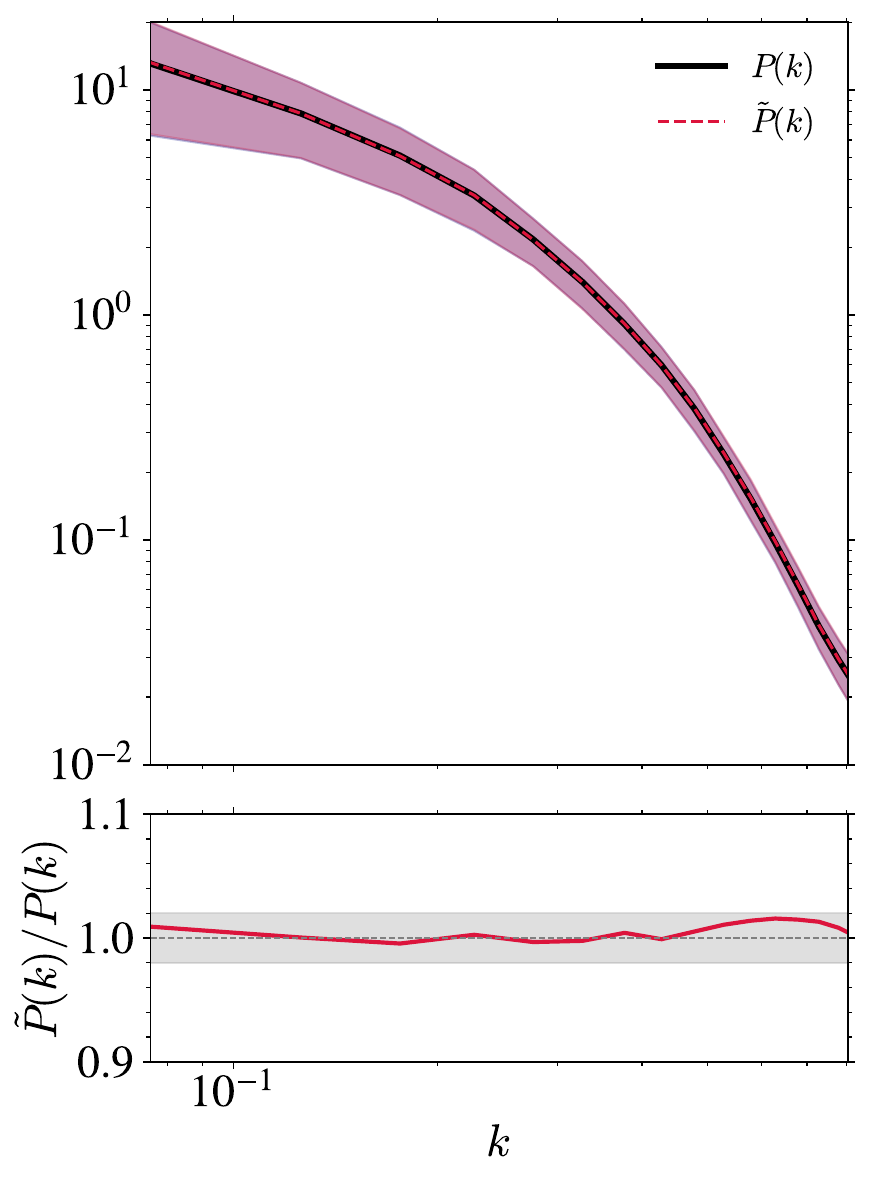}
        \caption{Left: $4\times4$ grids comprising $32\times32$ samples of both the true (left) and generated (right) DM log-density fields. Right: Mean matter power spectra comparison between true (solid black) and generated (dashed red) density fields. Shaded bands indicate the $1\sigma$ dispersion around the mean spectra in each case. The lower panel shows the ratio of mean amplitudes, using the true spectra as reference. The trained diffusion model reproduces the true mean power spectra of the density fields to within $\approx2\%$ across all scales, as indicated by the grey shaded band.}
        \label{fig:quijote_v_gen}
    \end{figure*}

   \subsection{Renormalisation group diffusion models}\label{RGDM_sec}

    Denoising diffusion models are a class of generative models that learn the score function of a data distribution, that is, the gradient of the log-probability density, by progressively transforming data into noise and then reversing this process. This framework is based on two coupled stochastic processes: a forward process that gradually corrupts data with Gaussian noise, and a reverse denoising process aimed at reconstructing samples from the original probability distribution by iteratively removing this noise.

    In this work, we adopted a renormalisation group diffusion model \cite[RGDM,][]{masuki2025generativediffusionmodelinverse}, which generalises standard denoising diffusion probabilistic models \cite[DDPMs,][]{ho2020denoisingdiffusionprobabilisticmodels} by introducing a scale-dependent noise schedule inspired by renormalisation group principles from statistical physics. Rather than injecting noise uniformly at all spatial scales, RGDMs progressively coarse-grain the data, enabling the model to learn how structures evolve across different physical scales.
    Given an input field $\mathbf{x}_0 \sim q(\mathbf{x}_0)$, with $\mathbf{x}_0 \in \mathbb{R}^d$, the forward process gradually transforms it into noise by successively suppressing small-scale information. 
    Formally, as in conventional DDPMs, it is defined by an interpolation between input and Gaussian samples at time $t$, namely
    \begin{equation} \label{eq:forward_RGDM}
    	x_{t,k} = \sqrt{\overline{\alpha}_{t,k}} x_{0, k} + \sqrt{\overline{\beta}_{t,k}} \epsilon_k~,
    \end{equation}
    where the noise schedule is defined through $\overline{\alpha}_{t,k} = K_t(k)$, a cutoff function going from one to zero, allowing the separation of slow and fast modes, and $\overline{\beta}_{t,k} = G_0(k) \left(1 - K_t(k)\right)$. With this formalism, the input data distribution $q(\mathbf{x}_0)$ at time $t=0$ converges to a coloured Gaussian field with covariance $G_0(k)$ at large $t$.
    The associated Langevin dynamics has the following conditional probabilities at any time $t$ with respect to an input field at time $t=0$,
   \begin{equation}
       q(\mathbf{x}_{t}|\mathbf{x}_{0}) = \prod_k\mathcal{N}\left(x_{t,k}; \sqrt{\Bar{\alpha}_{t}} x_{0,k}, \Bar{\beta}_{t,k}\mathbf{\mathbf{I}}\right)~,
   \end{equation}
   which allows for direct sampling of a noised input $\mathbf{x}_t$ without going through the entire dynamical process.

   Inverting this transport map is possible \citep{ANDERSON1982313} via a time-reversed stochastic process similar to Eq.~\eqref{eq:forward_RGDM}, provided the score function $\mathbf{s}(\mathbf{x}_t, t) = \nabla_{\mathbf{x}_t} \log q_t(\mathbf{x}_t)$ is known. In discrete form, the reverse update from $t$ to $t-1$ reads
   \begin{equation}
       \mathbf{x}_{t-1,k} = \frac{1}{\sqrt{\alpha_{t,k}}} \left[ \mathbf{x}_{t,k} + \beta_{t,k} \mathbf{s}(\mathbf{x}_t, t) \right] + \Theta(t-1) \sqrt{\beta_{t,k}} \epsilon_k~,
   \end{equation}
   where $\Theta(t)$ is the Heaviside function taking value one if $t>0$ and zero otherwise. In practice, the score function is unknown as it depends on $q(\mathbf{x}_0)$, so we learn it using a neural network approximation of the noise injected at time $t$ to a data $\mathbf{x}_0$, and minimise the following weighted $L_2$ loss
   \begin{equation}
       \mathcal{L}(\btheta) = \sum_{t=1}^T \lambda_t \mathbb{E}_{\mathbf{x}_0, \epsilon}\left[\lVert \epsilon - \epsilon_\btheta(\mathbf{x}_t, t) \rVert_2^2 \right]~,
       \label{rgdm_loss}
   \end{equation}
   where $\lambda_t$ are some weights uniformising the contribution of all times to the loss function. More details about the setup, the architecture, and the choice of hyperparameters can be found in Appendix~\ref{model_arc_setup_section}.

   \subsection{Training data}\label{sect_traininig_data}

    Our training data were extracted from $N$-body DM particle simulations in the \textsc{Quijote} suite, which currently comprises more than $82{,}000$ simulations spanning over $7{,}000$ cosmological models. In this work, we focused on simulations produced under the fiducial cosmology of \cite{Planck2018}, each evolving $512^{3}$ DM particles from their initial conditions at $z=127$ to $z=0$ in a simulation box of length $1~h^{-1}\mathrm{Gpc}$. For our training data, we used snapshots of the DM particle distribution at $z=0$, which were then interpolated onto a $256^{3}$ uniform cubic grid under the piecewise-cubic spline mass assignment scheme to obtain their matter density fields. 
    
    The diffusion model was trained exclusively on randomly sampled two-dimensional patches of $100{,}000$ log-transformed density fields extracted from $64$ simulation cubes. Due to projection effects, the morphology of walls is not well preserved in $2$D slices compared to other cosmic web environments. Consequently, their associated signals in our quantitative analyses are expected to be weaker. To avoid using contiguous slices, every third slice from the simulation cubes was extracted to construct the training dataset. Each patch has a physical depth of $\sim3.91~h^{-1}\mathrm{Mpc}$ and a side length of $125~h^{-1}\mathrm{Mpc}$, with a spatial resolution of $32\times32$ chosen for computational efficiency. Prior to training, each density field sample underwent random transformations to ensure a high training sample diversity, including random crops of size $32\times32$ from the full density field, random rotations in $90^{\circ}$ increments and random horizontal or vertical flips, each applied with a probability of $1/2$. To further reduce the dynamic range of the log-transformed density fields, all training samples were divided by $2$ prior to training.

    Figure \ref{fig:quijote_v_gen} (left) presents a visual comparison between the true density fields from \textsc{Quijote} and the generated samples after training. Visually, the generated samples clearly capture the prominent structural patterns of the cosmic web, reproducing the clustering of dense nodes interconnected by filaments, along with void regions that dominate most of the projected area of the $2$D samples.
    
    To quantify how well the model reproduces two-point statistics, we computed the mean matter power spectra of both the true and generated density fields over $1{,}000$ samples, as seen on the right panel of Fig.~\ref{fig:quijote_v_gen}. The shaded bands represent the $1\sigma$ dispersion of the mean matter power spectra for each case. We find excellent agreement between the mean amplitudes, with the trained model accurately reproducing the true power spectra to within $\approx 1\%$ across linear scales, and within $2\%$ towards more non-linear scales. As seen visually, the wide range of morphological structures captured across different cosmic web environments suggests that the trained model also encodes non-Gaussian information beyond two-point statistics. We show this quantitatively by computing the one-point probability density function (PDF) and the equilateral bispectrum of true and generated DM density fields (Fig.~\ref{fig:one_point_pdf_bispectra}). We find close agreement between the true and generated data in both cases, demonstrating that the model actually captures non-Gaussian information beyond two-point statistics. Considering this trained model, we qualitatively and quantitatively investigated the sensitivity of its self-attention maps to both the overall matter distribution and also individual cosmic web environments.
   
   \subsection{Self-attention maps}\label{sect_self_attn_maps}
   A typical diffusion model employs a neural network architecture composed of a contracting path of encoder blocks and an expanding path of decoder blocks, connected through a bottleneck. Each block consists of a series of convolutional neural networks \citep[CNNs,][]{6795724, lecun2015deep}. This symmetric encoder-decoder architecture, known as a U-Net \citep{ronneberger2015unetconvolutionalnetworksbiomedical}, enables the model to capture semantic information about the data structure across different spatial scales. To enhance this capability, self-attention maps \citep{vaswani2023attentionneed} are incorporated throughout several encoder and decoder blocks. Each region in the input density field is represented by a feature vector, which is linearly projected into three latent representations: queries $Q$, keys $K$, and values $V$. Queries represent the regions of interest, while keys define the regions with which those queries correlate. Finally, values encode latent information associated with each key. To obtain $Q$, $K$, and $V$, the linear projections are performed through learnable weight matrices, $W_{Q}$, $W_{K}$, and $W_{V}$, which are fine-tuned via training. The self-attention mechanism is implemented through the scaled dot-product function,
   \begin{equation}
   \text{Attention}(Q, K, V) = \text{softmax}\left(\frac{QK^{T}}{\sqrt{d_{k}}}\right)V~,
   \label{attention_function}
   \end{equation}
   which computes the dot-product between queries and keys, both of dimension $d_{k}$. The outputs of the softmax function act as attention weights which constitute the attention maps, encoding how strongly each pixel or spatial element in the input density field attends to all others, including itself. The collection of attention maps corresponding to all query positions defines an attention layer of a given spatial resolution. These layers $\mathcal{A}_{k}$ across the U-Net encoder-decoder blocks can be represented as four-dimensional tensors \citep{tian2024diffuseattendsegmentunsupervised}:
   \begin{equation}
       \mathcal{A}_{k} \in \left\{\mathcal{A}_{k} \in \mathbb{R}^{h_{k}\times w_{k}\times h_{k}\times w_{k}} \;\middle|\; k=1,...,n \right\}~,
       \label{attention_tensors}
   \end{equation}
    where $h_{k}\times w_{k}$ defines the spatial resolution of the attention maps in layer $k$ and $n$ is the total number of attention layers. Our model contains $21$ attention layers distributed throughout the U-Net, each being a four-dimensional tensor of size $r_{k}^{4}$, with spatial resolutions $r_{k}\in\{4, 8, 16, 32\}$.
    
    An attention map corresponding to a query position $(I, J)$ is given by $\mathcal{A}_{k}[I, J, :, :]\in \mathbb{R}^{h_{k}\times w_{k}}$, whose receptive field depends on the spatial resolution of the map. High-resolution maps have smaller receptive fields, meaning each query position covers a smaller region of the input density field. Such maps capture fine-grained, small-scale structures, whereas lower-resolution maps, having larger receptive fields, capture more global, large-scale features. This allows the model to capture structural patterns of the cosmic web on multiple scales. See Appendix \ref{sect_attention_map_extraction} for a detailed description of how attention maps are extracted from our trained model.

    Self-attention maps also exhibit characteristic properties that demonstrate how the model encodes spatial relationships \citep{tian2024diffuseattendsegmentunsupervised}. Key positions within maps corresponding to structures that share similar semantic and morphological characteristics with the query associated with that map tend to exhibit strong activations. For example, in an attention map corresponding to a query $(I, J)$, a neighbouring key position $(I + 1, J + 1)$ is likely to belong to the same structure as that query, resulting in a large activation value, $\mathcal{A}_{k}[I, J, I + 1, J + 1]$. This property is referred to as intra-attention similarity. Additionally, maps associated with query positions belonging to similar structures, for example, $\mathcal{A}_{k}[I, J, :, :]$ and $\mathcal{A}_{k}[I + 1, J + 1, :, :]$, also tend to have similar activation responses, known as inter-attention similarity. These properties are also observed in the extracted attention maps from our trained model (see Sect.~\ref{qualitative_analysis}).

   \subsection{Attention aggregation}

   To quantitatively assess how well the attention maps in our trained diffusion model capture prominent cosmic web structures, we adopted the concept of attention aggregation, introduced by \cite{tian2024diffuseattendsegmentunsupervised}. This involves aggregating attention layers of varying spatial resolutions into a single, high-resolution ($r=32$) attention layer that provides a multi-scale representation of cosmic web structures. For our analysis, we used $9$ out of $21$ attention tensors in our model for aggregation, as discussed in Appendix \ref{sect_attention_map_extraction}. The different resolutions of these attention tensors allow individual attention maps to capture cosmic web patterns at various scales determined by their receptive fields. Each map within this aggregated tensor thus combines both small- and large-scale structural patterns of the cosmic web.

   To achieve this, we first upsampled the key dimensions (the last two axes of the $4$D tensors) of all lower-resolution attention tensors using bilinear interpolation, such that each query position has an associated $32\times32$ attention map. For each attention tensor in Eq.~(\ref{attention_tensors}),
   \begin{equation}
       \tilde{\mathcal{A}}_{k} = \text{Bilinear-upsample}(\mathcal{A}_{k})\in \mathbb{R}^{h_{k}\times w_{k}\times32\times32}~.
       \label{upsample_bilinear}
   \end{equation}
   We then computed the final aggregated attention as a weighted sum of these upsampled tensors, where the weight assigned to each tensor is proportional to its spatial resolution. Thus, in the resulting aggregated attention $\mathcal{A}_{f}\in\mathbb{R}^{32^{4}}$, each map associated with a given aggregated query position receives weighted contributions from the maps across all attention tensors whose query positions spatially overlap with the region covered by that aggregated query. Following \cite{tian2024diffuseattendsegmentunsupervised}, the aggregated attention tensor is given by
   \begin{equation}
       \mathcal{A}_{f}[I, J, :, :] = \sum_{k\in \mathcal{K}}\tilde{\mathcal{A}}_{k}\left[\frac{I}{\delta_{k}}, \frac{J}{\delta_{k}}, :, :\right]R_{k}~,
       \label{aggregated_attention_tensor}
   \end{equation}
   where $\delta_{k} = 32 / w_{k}$, $\sum_{k}R_{k}=1$, and the divisions in both Eq.~(\ref{aggregated_attention_tensor}) and in $\delta_{k}$ rescale $(I, J)$ down to the nearest grid indices. The aggregation importance ratio $R_{k}$ is defined as 
   \begin{equation}
   R_{k}=\frac{w_{k}^{\alpha}}{\sum_{k}w_{k}^{\alpha}}~,
   \label{aggregation_importance_ratio}
   \end{equation}
   which determines the relative contribution of maps of each attention tensor, with $\alpha$ controlling the strength of this resolution-dependent weighting. Following \cite{tian2024diffuseattendsegmentunsupervised}, we adopted $\alpha=1$ as our fiducial choice, such that higher-resolution attention tensors are assigned proportionally greater weight. The robustness of this choice is verified in Fig.~\ref{fig:cross_ps_ps_agg_attn_multiple_alphas_fracs} (left panel). Each map in the final aggregated attention tensor is further normalised to ensure that it represents a valid probability distribution. With this inclusion of aggregated attention, we performed both qualitative and quantitative analyses of multiple attention tensors to evaluate how effectively they capture distinct cosmic web environments.

   \subsection{T-Web classification}

   Assessing the ability of self-attention maps to capture cosmic web structures requires a reference segmentation of the matter density field, where each grid cell is classified under one of the main cosmic web environments: voids, walls, filaments, and nodes. We adopted this segmentation as our ground truth, obtained using the T-Web algorithm \citep{Hahn_2007}, following the implementation described in \cite{Bonnaire_2022}.

    \begin{figure}[htbp]
        \centering
        \includegraphics[width=\columnwidth]{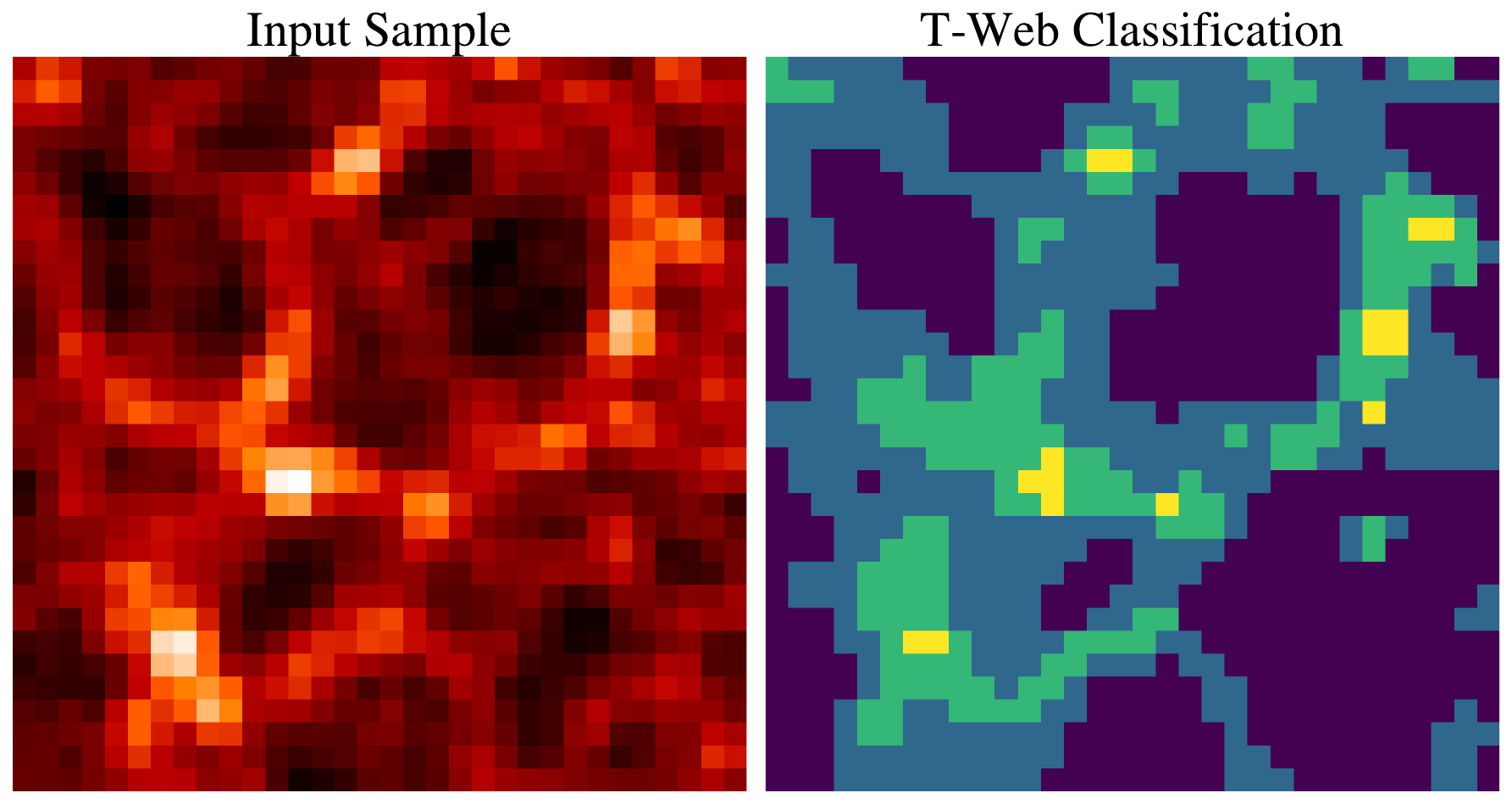}
        \caption{Left: $32\times32$ patch of a log-transformed density field from \textsc{Quijote}. Right: The corresponding T-Web segmentation mask, highlighting the four cosmic web environments in order of increasing pixel intensity: voids, walls, filaments, and nodes.}
        \label{fig:quijote_v_T_Web}
    \end{figure}
   
   This segmentation method classifies each grid cell based on the eigenvalues of the tidal tensor, defined as the Hessian of the gravitational potential derived from the matter density field. The number of eigenvalues exceeding a chosen threshold determines the environment assigned to each grid cell. Figure~\ref{fig:quijote_v_T_Web} illustrates a $2$D slice of a log-transformed matter density field from \textsc{Quijote} and its corresponding T-Web classification.

    \begin{figure}[htbp]
        \centering
        \includegraphics[width=\columnwidth]{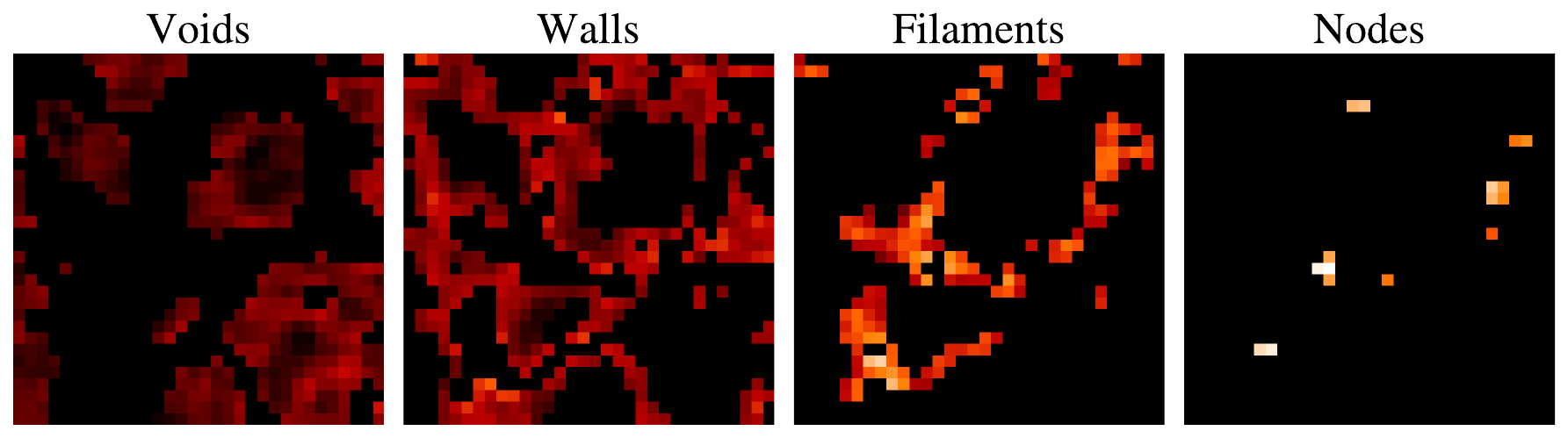}
        \caption{Four cosmic web environment density fields of the input sample in Fig. {\ref{fig:quijote_v_T_Web}}, classified by T-Web and scaled logarithmically.}
        \label{fig:al_4_comp_fields}
    \end{figure}

   Based on T-Web classification, the full three-dimensional matter density field can be decomposed into four cosmic web environment density fields $\{\delta_{v}, \delta_{w}, \delta_{f}, \delta_{n}\}$ (see Fig.~\ref{fig:al_4_comp_fields}) by retaining cells assigned to a given environment and setting all remaining cells to the minimum log-density value. The T-Web environment masks are also used to group attention maps according to the environment associated with their query pixel positions. Several statistical estimators are then used to quantitatively assess how effectively these environment-associated attention maps capture cosmic web environments in the DM density field.

\section{Statistical estimators}\label{sec_stat_estimators}

    In this section, we present three statistical estimators employed for the quantitative analysis of attention maps, and the motivation behind their choices.
    
   \subsection{Dice coefficient}\label{section_dice_coeff}

    We utilised the Dice coefficient \citep{https://doi.org/10.2307/1932409} to measure the degree of spatial overlap between a log-transformed density field and the environment-associated attention maps produced by the trained model for that input field. We analysed this overlap by introducing an intensity threshold parameter $\tau \in [0, 1]$, which defines the fraction $\tau$ of pixels with the lowest intensities to be retained, producing binary masks that preserve values equal to or below the corresponding threshold. We expressed this statistic as
   \begin{equation}
       \textrm{Dice}(\delta_{m}, A_{\beta}, \tau) = \frac{2\lvert \delta_{m}(\tau)\cap A_{\beta}(\tau)\rvert}{\lvert \delta_{m}(\tau)\rvert + \lvert A_{\beta}(\tau)\rvert}~,
       \label{dice_coefficient}
   \end{equation}
   where $\delta_{m}(\tau)$ and $A_{\beta}(\tau)$ are the binary masks of the log-transformed density field and the attention map associated with environment $\beta\in\{v, w, f, n\}$, respectively. Evaluating the Dice coefficient in this way provides a global measure of how well attention maps capture structures across different density regimes.
   
   \subsection{Cross-power spectra}\label{section_cross_power_spectra}

   To evaluate the correspondence between matter density fields and their attention maps across spatial scales, we computed the cross-power spectrum, defined as
   \begin{equation}
        P_{\delta, A_{\beta}}(k) = \left\langle \tilde{\delta}(\mathbf{k}) \tilde{A_{\beta}}^{*}(\mathbf{k}) \right\rangle_{k}~,
        \label{cross_power}
    \end{equation}
    where $\tilde{\delta}(\mathbf{k})$ and $\tilde{A_{\beta}}(\mathbf{k})$ represent the Fourier transforms of a log-transformed density field $\delta$ and an environment-associated attention map $A_{\beta}\in\{A_{v}, A_{w}, A_{f}, A_{n}\}$, respectively. Here, $\delta$ can be either the matter density field $\delta_{m}$, or an environment density field  $\delta_{\alpha}\in\{\delta_{v}, \delta_{w}, \delta_{f}, \delta_{n}\}$. The average is taken over all modes with the same wavenumber $k$. For our analysis, we computed the following cross-power spectra:
    \begin{itemize}
        \item $P_{\delta_{m}, A_{\beta}}(k)$: cross-power between the matter density field $\delta_{m}$ and environment-associated attention maps, to assess the sensitivity of such maps to different spatial scales in the overall density field.
        \item $P_{\delta_{\alpha}, A_{\beta}}(k)$: cross-power between environment density fields $\delta_{\alpha}$ and environment-associated maps, to quantify how well these maps capture the structures of individual cosmic web environments.
    \end{itemize}

   \subsection{Cross-environment attention distributions}\label{sect_cross_comp}

   Another effective way to quantify how attention maps respond to different cosmic web environments is by analysing the distributions of their key values. For all maps whose queries are associated with a given environment, we extracted their key values and classified them according to the environments that their key positions belong to, again using T-Web masks as reference. 
   
   This approach provides us with a set of key distributions grouped by the environments associated with both their query and key positions. By comparing the PDFs of these cross-environment key distributions, we can assess how strongly attention maps of one environment respond to another environment, for example, void attention maps may exhibit higher key values in void-like (underdense) regions than in node (high-density) regions. This provides insight into how the model's attention maps encode latent representations of cosmic web structures and the relationships between their environments.

    \begin{figure*}[htbp]
        \centering
        \includegraphics[width=0.33\textwidth]{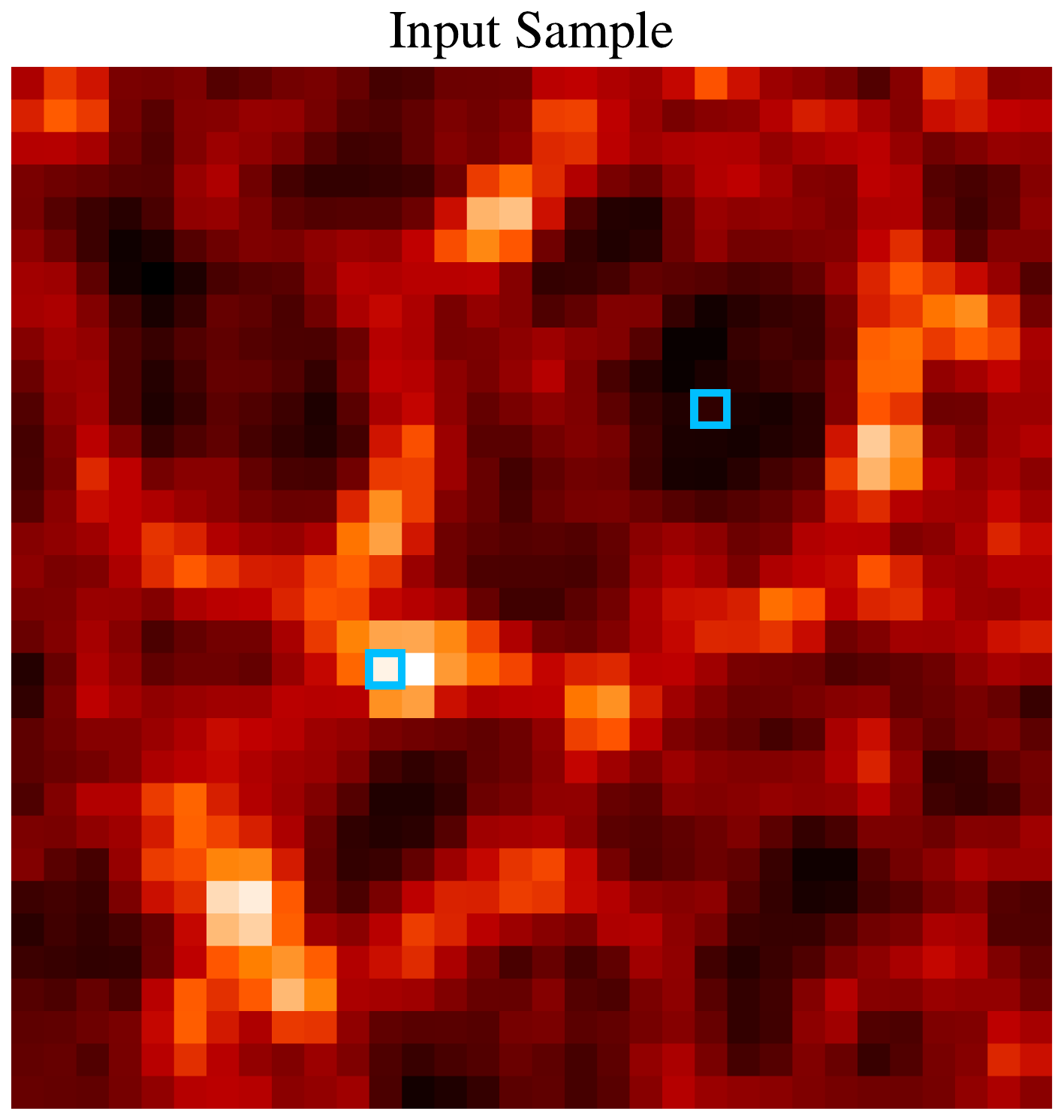}
        \includegraphics[width=0.33\textwidth]{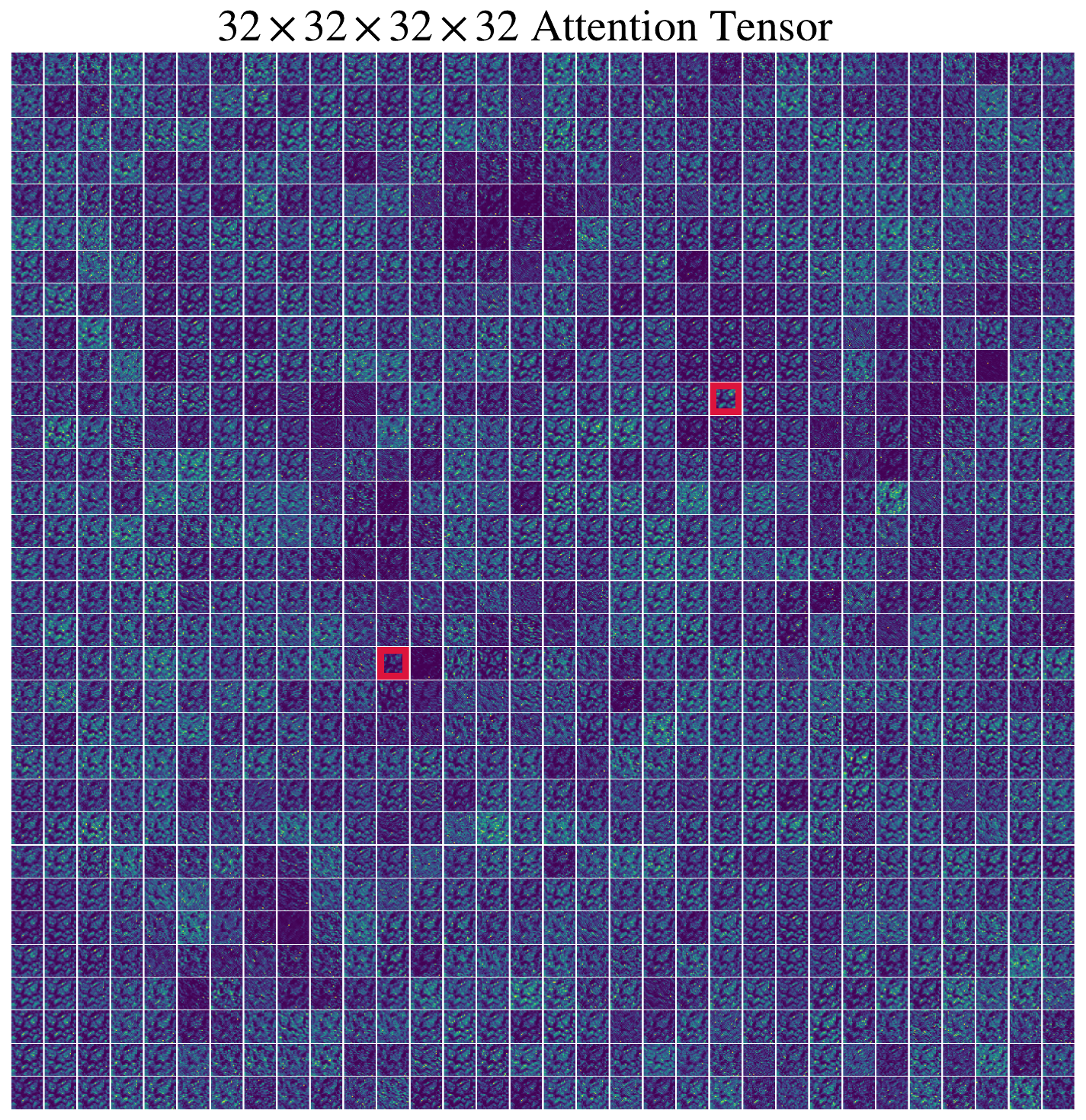}
        \includegraphics[width=0.33\textwidth]{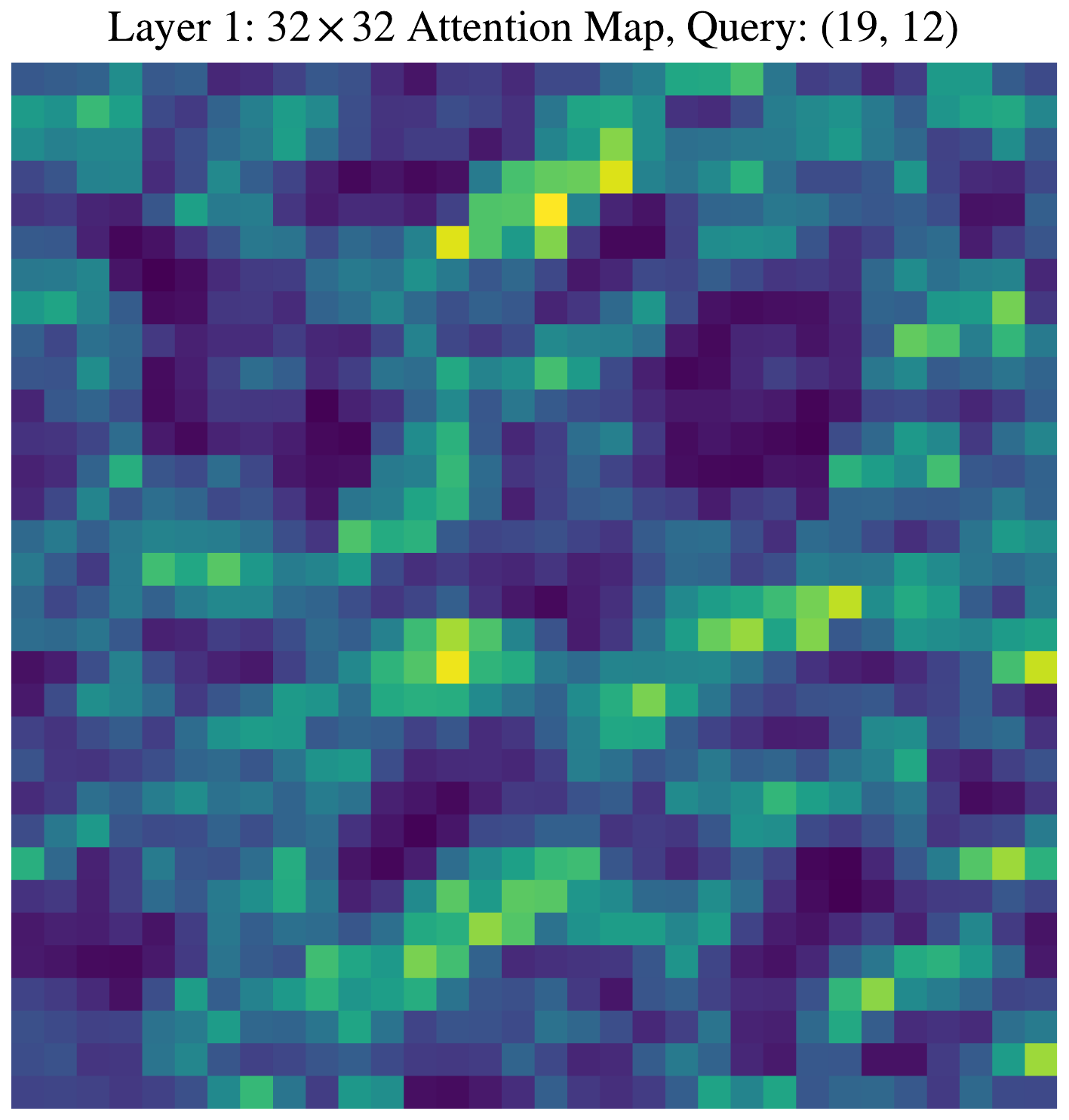}
        \\[0.4em]
        \includegraphics[width=0.33\textwidth]{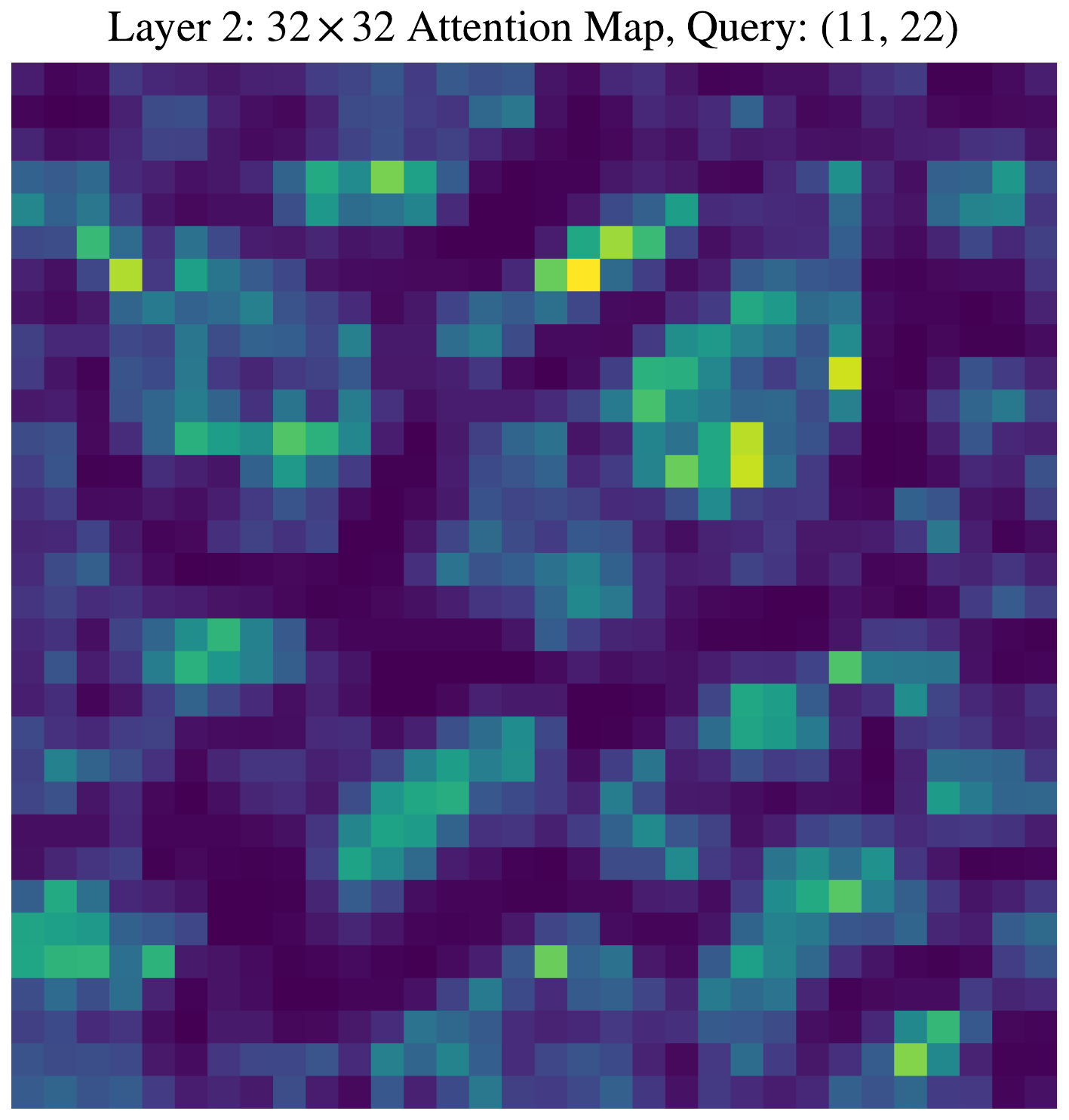}
        \includegraphics[width=0.33\textwidth]{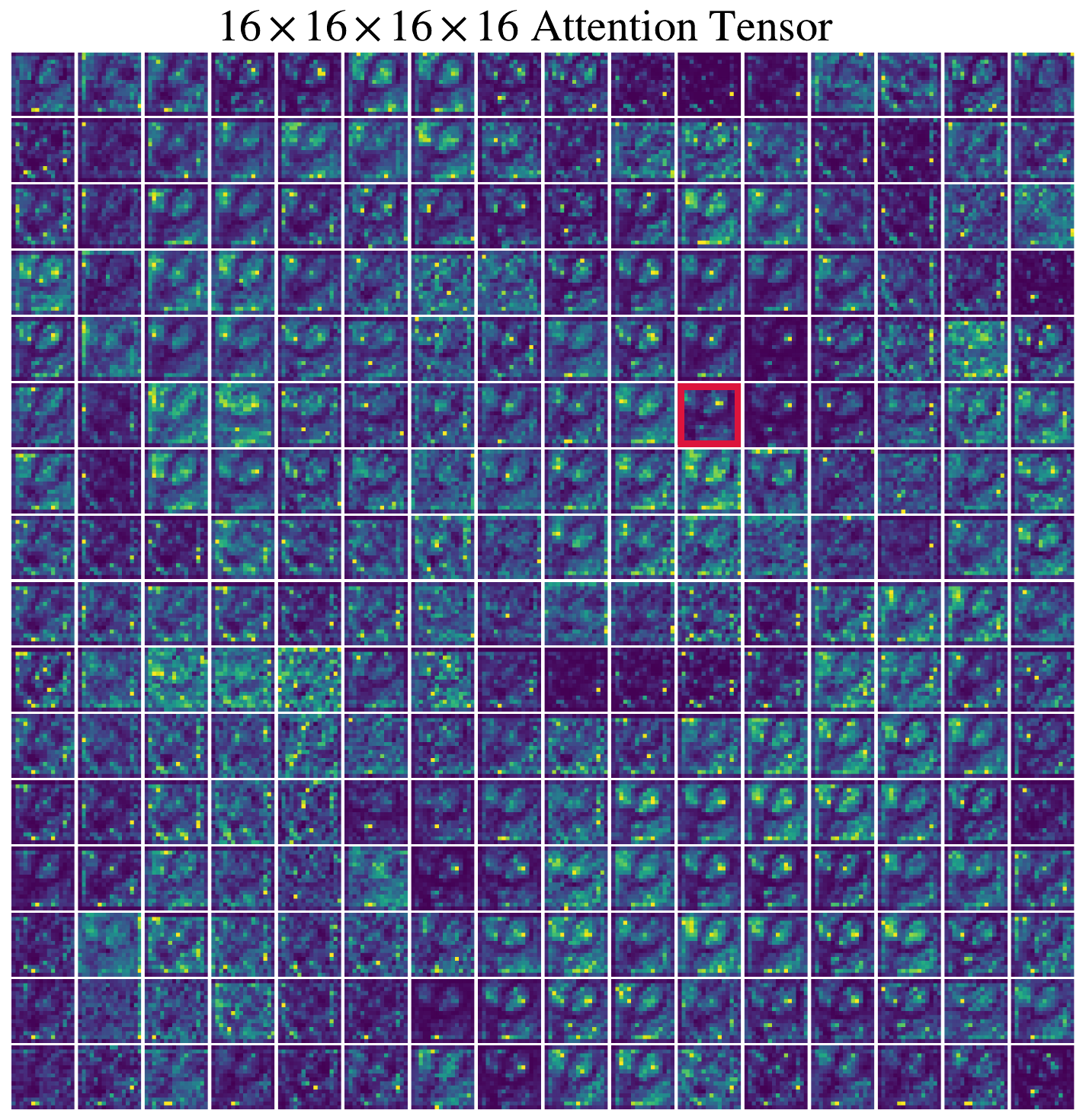}
        \includegraphics[width=0.33\textwidth]{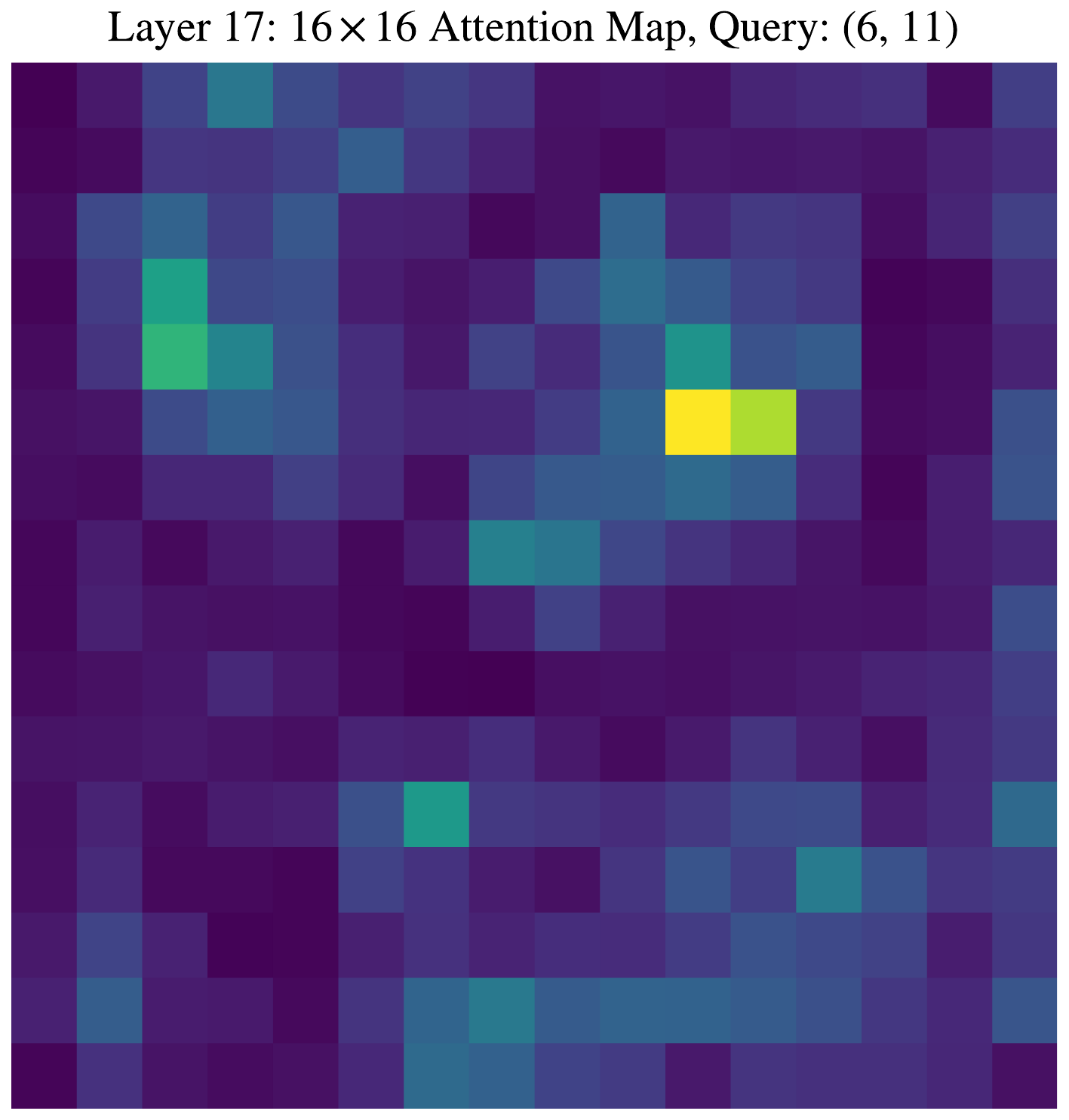}
        \caption{Illustration of multi-resolution attention maps associated with a $2$D slice of a matter density field from \textsc{Quijote}. First row (from left to right): An input density field, a $32\times32\times32\times32$ attention tensor from the trained diffusion model containing an attention map for each spatial pixel of the input, and an attention map from layer $1$ associated with the node query position $(19, 12)$ as classified by T-Web. The map exhibits strong activations in overdense regions, highlighting filaments and nodes. Second row (from left to right): An attention map from layer $2$ corresponding to the void query position $(11, 22)$, a $16\times16\times16\times16$ attention tensor, and a $16\times16$ map from layer $17$ corresponding to a void query position overlapping with the $(11, 22)$ pixel in the input field. Both maps respond strongly to underdense, void-like regions, with the high-resolution case capturing small-scale, finer structures, while the lower-resolution map captures larger-scale features due to its broader receptive field. Each map in the attention tensors is shown with its own scale for visual clarity.}
        \label{fig:qualitative_plots}
    \end{figure*}

\section{Results}\label{sec_results}

    From our trained model, we extracted self-attention maps from multiple attention tensors to qualitatively and quantitatively assess their ability to capture prominent cosmic web structures, given an input DM density field from the \textsc{Quijote} suite.

    \subsection{Qualitative analysis}\label{qualitative_analysis}
    
    The aforementioned properties of self-attention maps that demonstrate how spatial relationships are encoded (see Sect.~\ref{sect_self_attn_maps}) are also observed in our extracted maps from the trained model, which we first assessed through visual inspection. Examples of this are shown in Fig.~\ref{fig:qualitative_plots}, where the first row illustrates a given input density field, an example of the highest-resolution attention tensor $\mathcal{A}_{k}\in{\mathbb{R}^{32}}^{4}$, and an attention map of that resolution, respectively. The corresponding pairs of blue and red squares in the input sample and attention tensor highlight two query positions: $(19, 12)$ which belongs to a node environment, and $(11, 22)$ which belongs to a void environment. 
    
    The node-associated map in the first row clearly exhibits strong activation of key values in regions dominated by overdense structures, capturing the overall filamentary pattern along with node regions. In contrast, the void-associated map, displayed in the first column of the second row, shows a strong response in key positions clearly dominated by underdense regions, capturing void-like environments. This behaviour of strong key activations belonging to structures similar to its query position indicates intra-attention similarity. Maps associated to queries neighbouring the void positions of our interest also show similar activation responses, as seen in the attention tensors of both the first and second row in Fig.~\ref{fig:qualitative_plots}. Thus, the activation of keys corresponding to similar filamentary or void-like structures, along with similar activations across maps whose queries belong to the same cosmic web environment, demonstrate both intra-attention and inter-attention similarity.
    
    The scale of structures captured by these maps is determined by the size of their receptive fields in the input density field (see Sect.~\ref{sect_self_attn_maps}). To illustrate this, we examined a lower-resolution attention tensor and one of its maps, shown in the second row of Fig.~\ref{fig:qualitative_plots}. Maps at this resolution have a larger receptive field, now covering a spatial region of a $2\times2$ grid of pixels in the input field, instead of the $1$-$1$ pixel correspondence of the higher-resolution maps. Upon visual inspection, we observe that the map associated with the $(6, 11)$ query position in the tensor, which covers a larger area of the same void region previously considered, captures larger-scale patterns of underdense regions, compared to the smaller-scale, fine-grained features present in the higher-resolution maps.

    \subsection{Quantitative analysis}\label{sect_quantitative}

    \begin{figure*}[htbp]
        \centering
        \includegraphics[width=\textwidth]{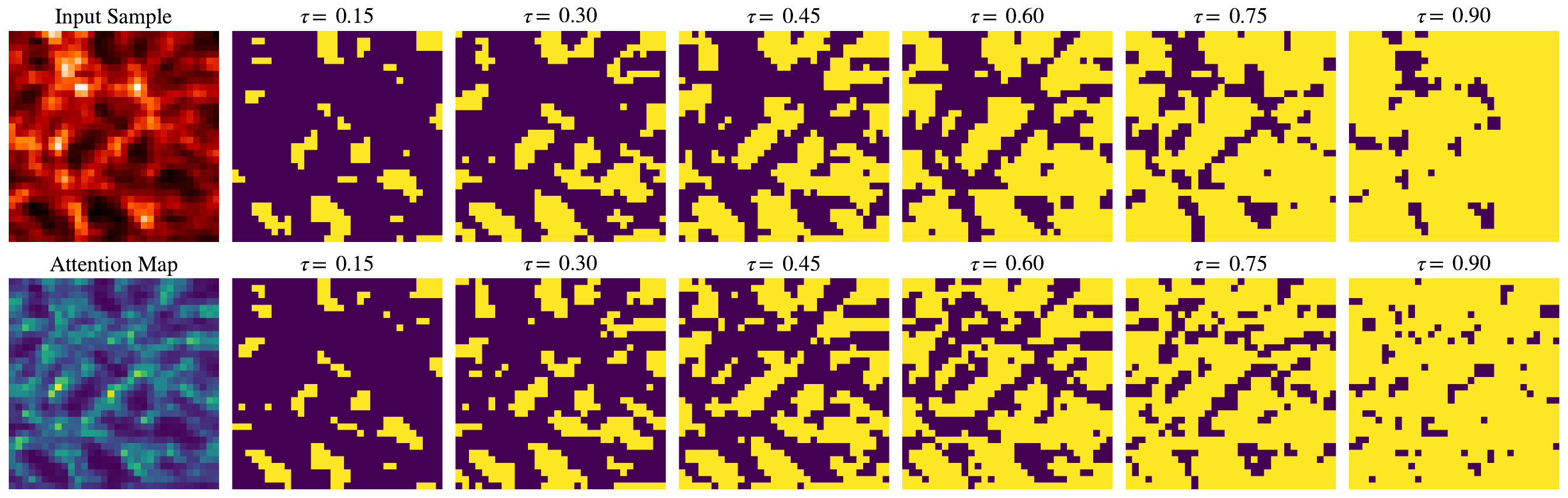}
        \\[0.4em]
        \includegraphics[width=0.246\textwidth]{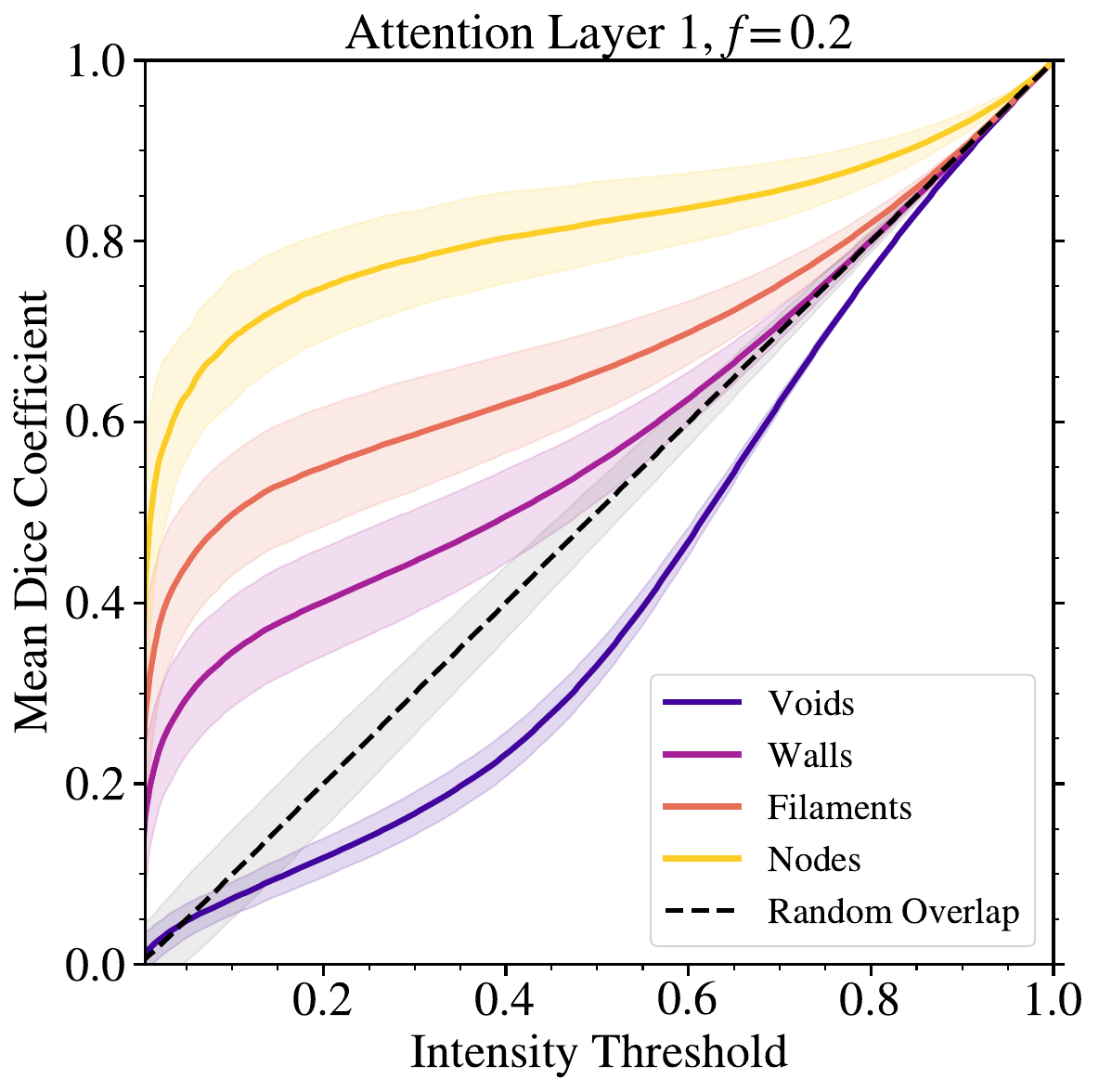}
        \includegraphics[width=0.246\textwidth]{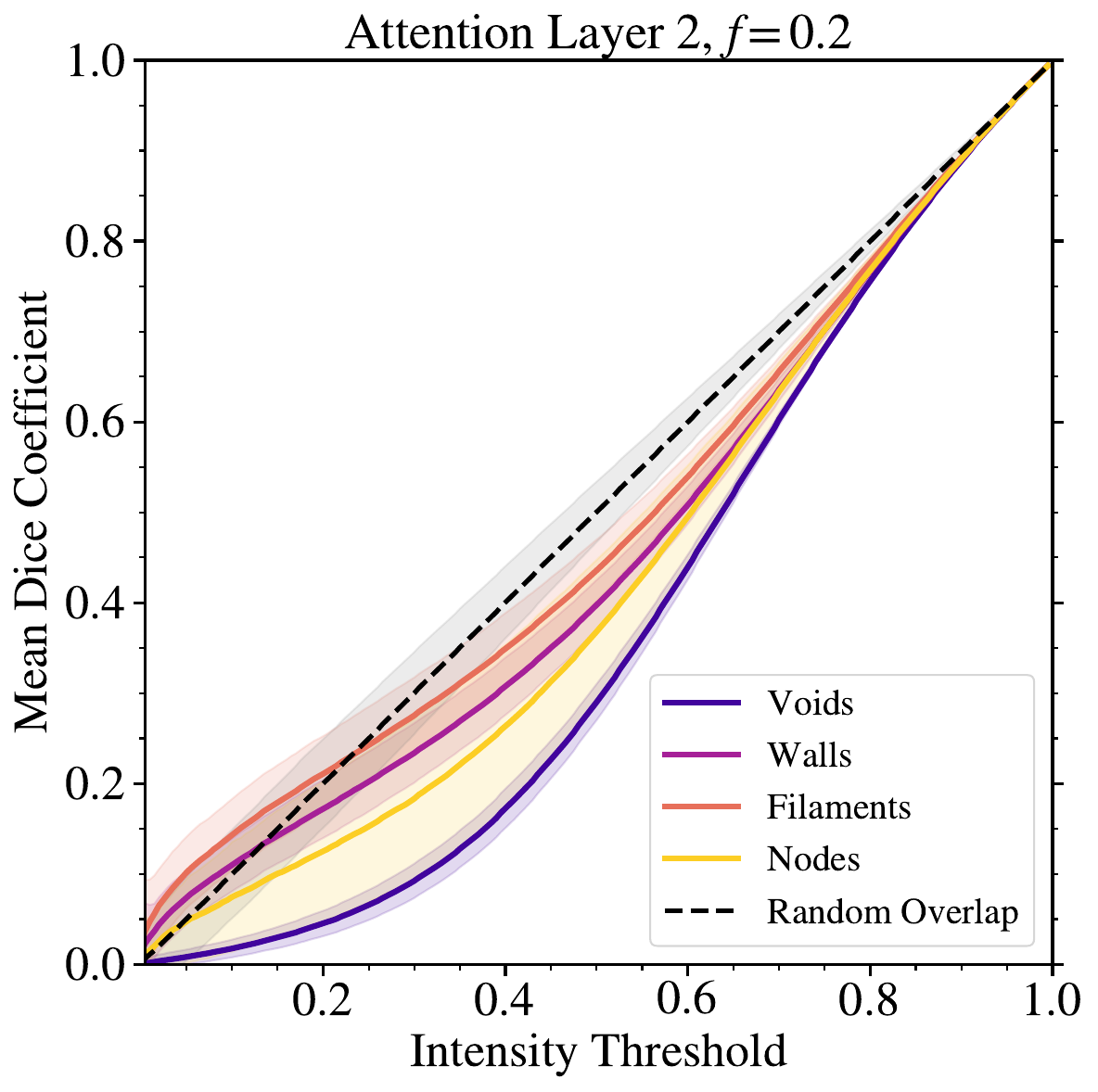}
        \includegraphics[width=0.246\textwidth]{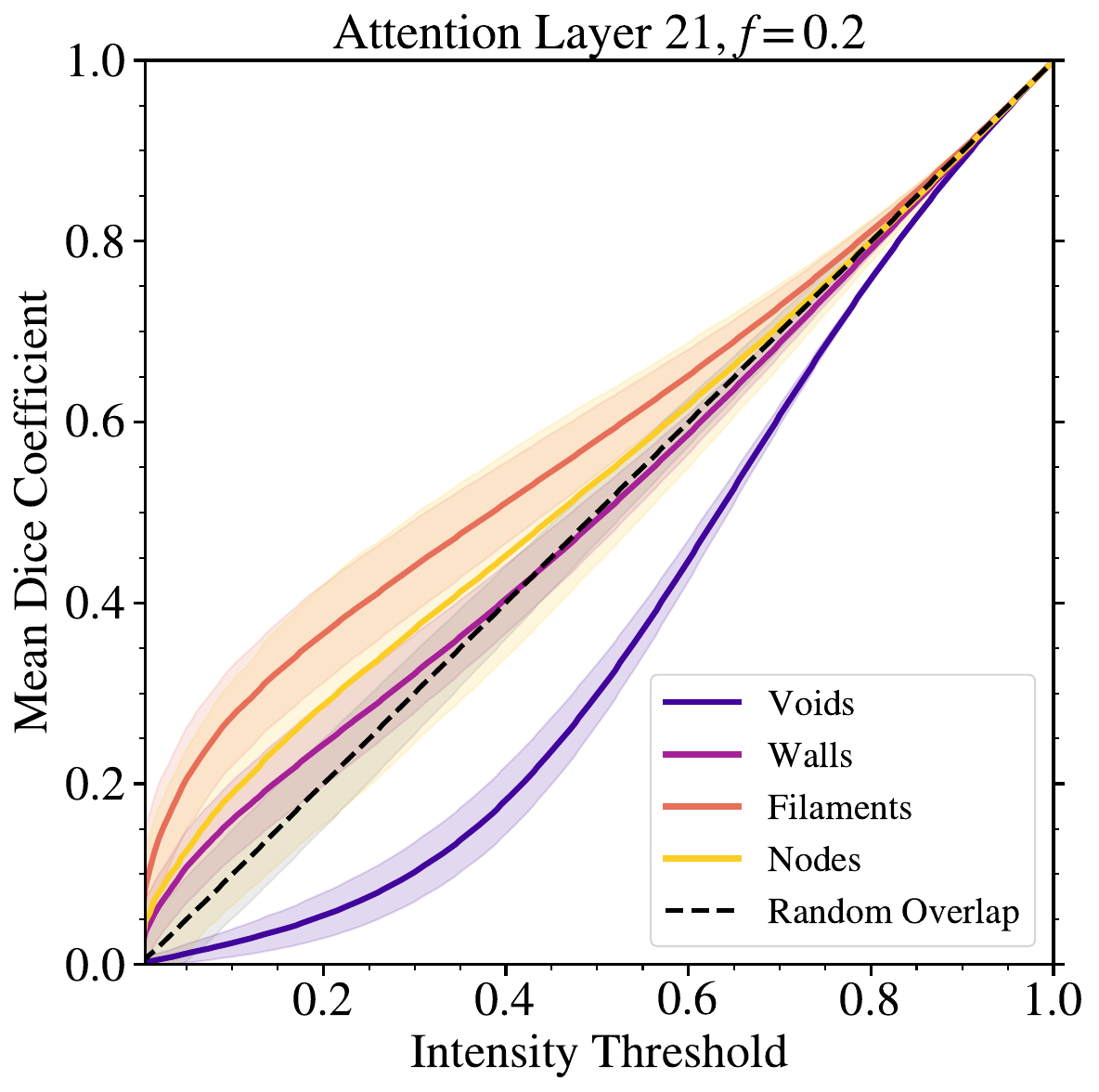}
        \includegraphics[width=0.246\textwidth]{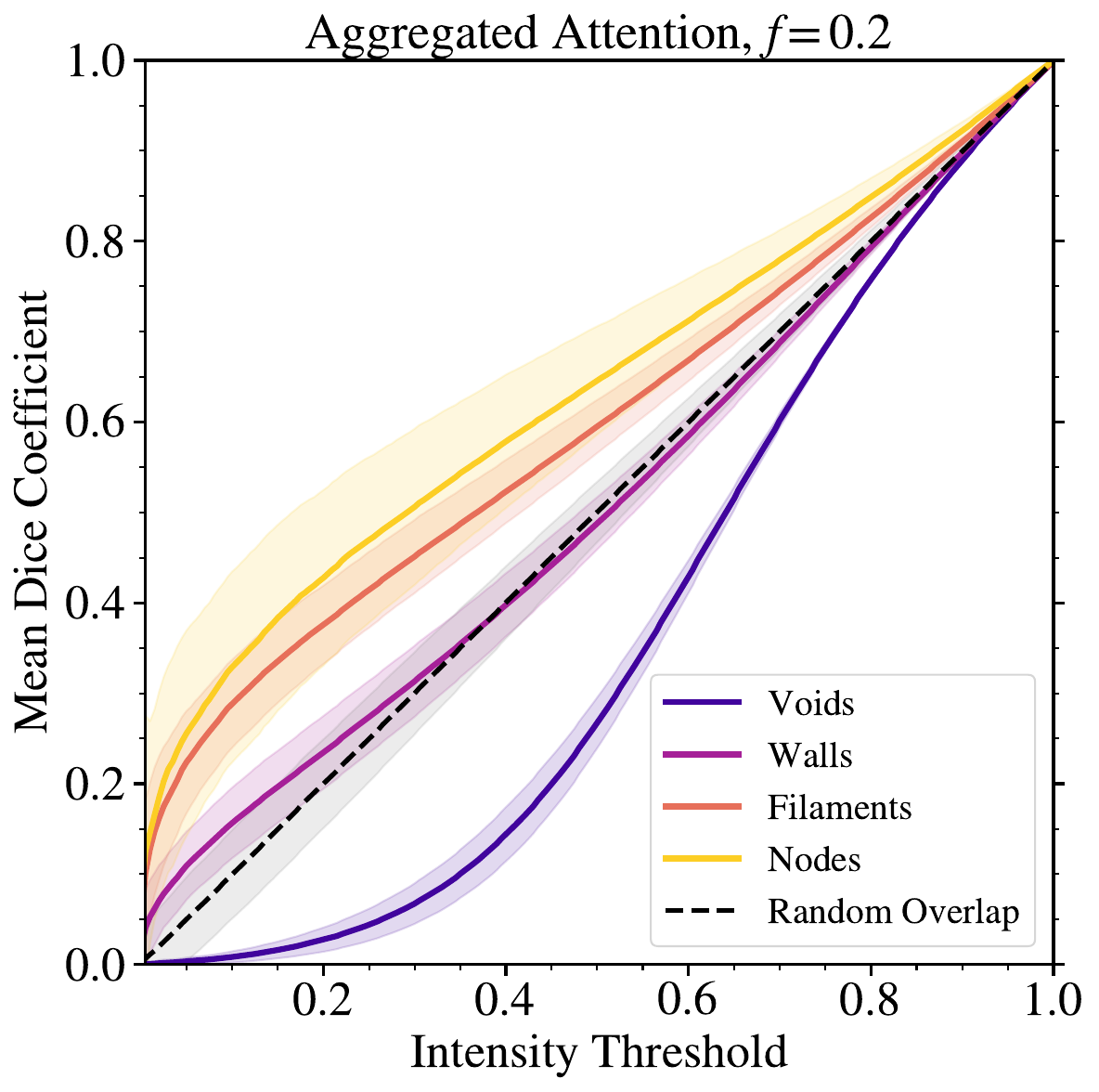}
        \caption{Top: Binary maps of an input sample and an associated attention map, produced under varying pixel intensity thresholds. Bottom: Dice coefficient curves for attention layers $1$, $2$, $21$, and the aggregated attention, quantifying the structural overlap between attention maps associated with different cosmic web environments and their corresponding matter density fields across multiple thresholds. Shaded bands show the $1\sigma$ dispersion of the mean Dice coefficient across $200$ samples at each threshold. Layers $1$ and $2$ predominantly show strong responses to overdense and underdense structures, respectively, while layer $21$ shows a weaker sensitivity to both. The aggregated attention displays strong correlations with both overdense and underdense regions.}
        \label{fig:dice_coefficients}
    \end{figure*}
    
    We quantified the information encoded in the attention maps through statistical analyses using various estimators. To mitigate noise and projection effects arising from the use of $2$D slices and limited spatial resolution, we applied a thresholding procedure to the density fields, controlled by a parameter $f$ set to $0.2$, which retains the most characteristic regions of each cosmic web environment (see Sect.~\ref{sec_discussion} for more details).

    \subsubsection{Dice coefficient analysis}
    We analysed the spatial overlap between cosmic web structures in matter density fields and their corresponding attention maps associated with different cosmic web environments. The top two rows of Fig.~{\ref{fig:dice_coefficients}} show an input sample and an associated attention map, along with their binary maps produced at various pixel intensity thresholds, from which we can observe an important overlap between underdense structures at low thresholds $\tau$. For the Dice analysis, we considered $200$ density field samples. For each sample, we computed the Dice coefficients across all thresholds $\tau$ between each density field and its associated attention maps for a given environment. The resulting Dice curves were first averaged over all maps per environment to obtain a mean Dice curve per environment, which were then averaged over all samples to produce the global Dice curve for each environment. The resulting Dice curves per environment across multiple high-resolution layers, including the aggregated attention, are shown in the bottom row of Fig.~{\ref{fig:dice_coefficients}}. The shaded bands represent the $1\sigma$ dispersion of the mean Dice coefficient across $200$ samples at each threshold. To assess the significance of structural overlap between density fields and environment-associated attention maps, we compared the Dice curves against the diagonal dashed line, representing the expected level of random overlap between uncorrelated fields. 
    
    The filament and node environment Dice curves in attention layer $1$ significantly exceed the random overlap reference line across all intensity thresholds, indicating strong correlations with overdense structures, particularly in node maps. In contrast, the wall environment curve lies close to the reference, while the void curve lies well below it, exhibiting a good sensitivity to underdense regions. This is consistent with the patterns observed in Fig.~\ref{fig:qualitative_plots}, where layer $1$ shows strong activations primarily at filamentary and node regions. 
    
    By contrast, the layer $2$ Dice curves fall well below the reference for all environments. This indicates an inverse structural relationship with the density field, where maps in this layer, particularly void maps, are predominantly anti-correlated and highlight underdense, void-like regions. This is also consistent with the attention map from layer $2$ seen in Fig.~\ref{fig:qualitative_plots}.
    
    Layer $21$ demonstrates a sensitivity to both underdense and overdense structures, with the void curve below the reference and the filament and node curves above it. The structural overlap is notably weaker compared to layer $1$, as it displays more noise likely due to it belonging to a decoder block. 
    
    Finally, the aggregated attention, which combines structural information captured across multiple layers and resolutions, exhibits clear separations of the curves between overdense and underdense environment maps. The node and filament Dice curves significantly exceed the reference, while the void curve falls well below it. Additionally, the remaining high-resolution layers, $19$ and $20$ (see Fig.~\ref{fig:dice_19_20}), also belonging to decoder blocks, display significantly higher noise levels in their maps, and thus their results are not very informative. Overall, the Dice analysis demonstrates that different attention layers capture distinct cosmic web environments, with certain layers strongly correlated with overdense structures and others preferentially highlighting underdense environments in the matter density field.

    \subsubsection{Cross-power spectra analysis}\label{sect_cross_power_analysis}
\begin{figure*}[htbp]
        \centering
        \includegraphics[width=\textwidth]{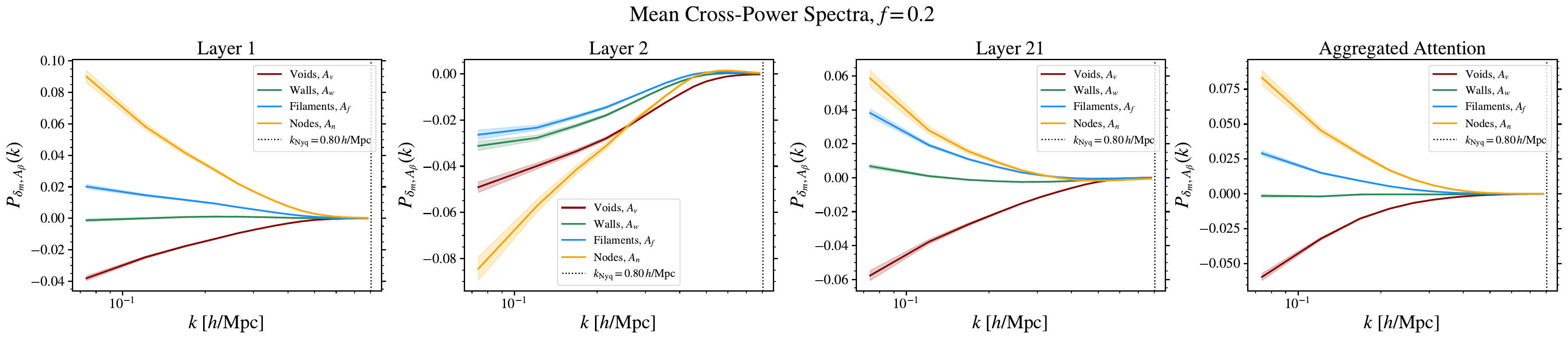}
        \caption{Cross-power spectra between log-transformed DM density fields and their corresponding $32\times32$ environment-associated attention maps from the trained diffusion model for attention layers $1$, $2$, $21$, and the aggregated attention. Dotted vertical lines indicate the Nyquist frequency, and the shaded bands represent $\pm3\mathrm{SEM}$ over $2000$ density field samples. Layers $1$ and $2$ exhibit strong positive correlations between node and filament maps with overdense structures, and anti-correlations between all environment maps and underdense structures, respectively. Layer $21$ displays a good sensitivity to both overdense and underdense structures, while the aggregated attention shows an even greater sensitivity to nodes and voids. Walls generally show weak signals in all four cases.}
        \label{fig:cross_power}
    \end{figure*}

\begin{figure}[htbp]
        \centering
        \includegraphics[width=\columnwidth]{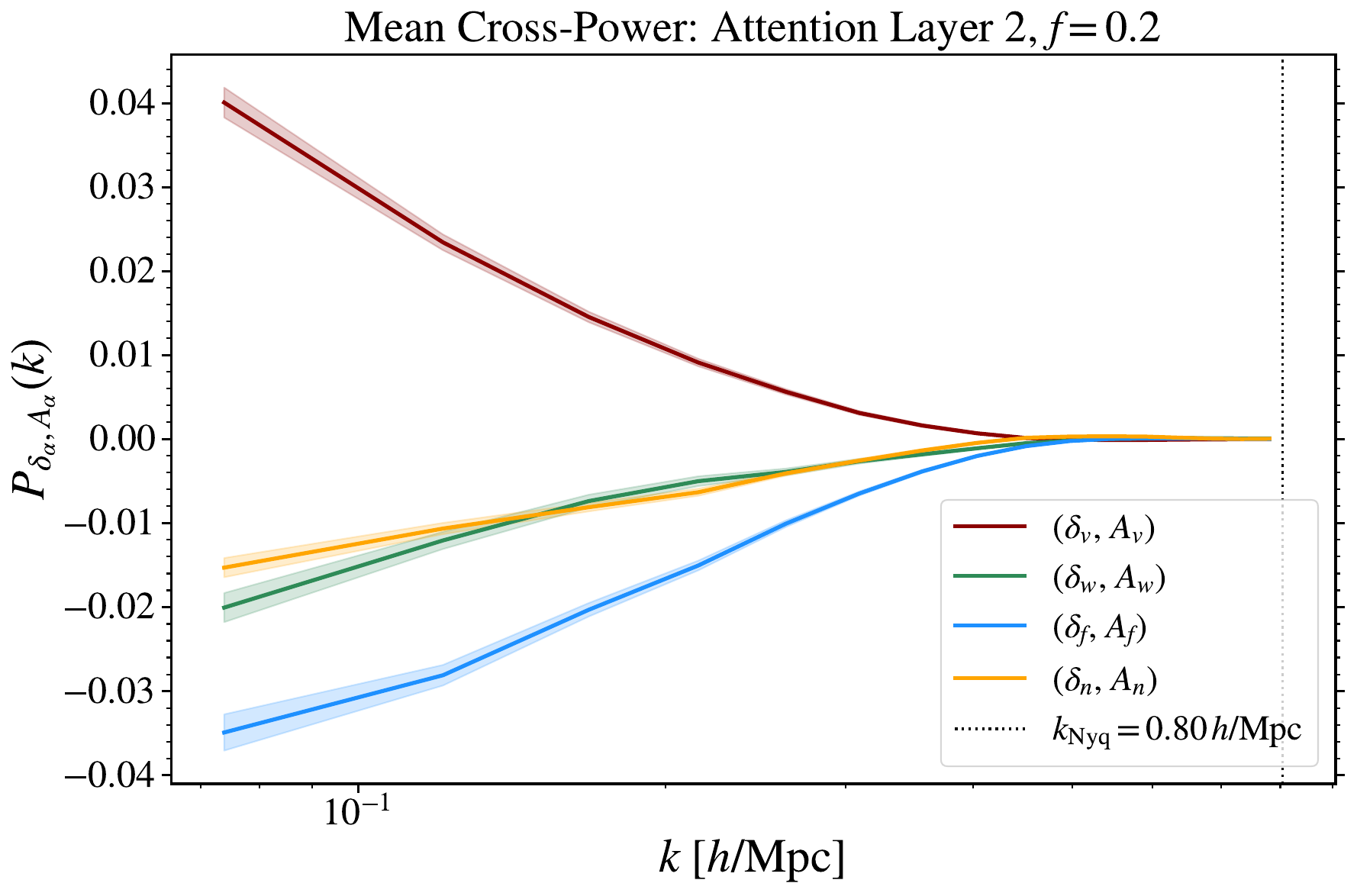}
        \caption{Cross-power spectra between matching environment pairs of cosmic web environment density fields and attention maps in attention layer $2$. Shaded bands represent $\pm3\mathrm{SEM}$ of $2000$ density field samples.}
        \label{fig:cross_comp_cross_ps_layer_2_same_comb}
    \end{figure}

    To quantify how attention maps encode structural information across spatial scales, we computed the cross-power spectra (see Sect.~\ref{section_cross_power_spectra}) between attention maps and the matter density field $\delta_{m}$, as well as their cosmic web environment density fields $\delta_{\alpha}\in\{\delta_{v}, \delta_{w}, \delta_{f}, \delta_{n}\}$ in Fourier space. The results for the same $32\times32$ attention layers analysed in the Dice study are shown in Fig.~\ref{fig:cross_power} for the matter density field and in Fig.~$\ref{fig:cross_comp_cross_ps_layer_2_same_comb}$ for the environment density fields. For each cosmic web environment, we computed the cross-power spectra between $2000$ input density field samples and their associated attention maps. The shaded bands show $\pm 3$ times the standard error of the mean (SEM) cross-power spectra for each environment within a given layer. This enables us to identify the spatial scales at which attention maps exhibit the strongest correlations with the density fields and, in turn, the scales that are most informative to the diffusion model. Details of the preprocessing, particularly for the analysis of lower-resolution maps, are provided in Appendix~\ref{sect_preprocessing_cross_power}.
    
    \paragraph*{Cross-power with matter density fields.}
    Across the considered attention layers, the strongest correlations and anti-correlations are observed at intermediate to large spatial scales, with the signal gradually decaying towards smaller scales, where the $\pm3$SEM bands are narrowest due to the larger number of Fourier modes contributing at higher $k$. For $32\times32$ attention maps, significant correlations appear at scales of $\sim21$ - $125~h^{-1}\mathrm{Mpc}$ ($\sim0.05$ - $0.3~h\mathrm{Mpc}^{-1}$), with the Nyquist frequency corresponding to a physical scale of $\sim7.81~h^{-1}\mathrm{Mpc}$.

    The scales at which correlations are captured by the trained model are directly related to the length scales spanned by the input density field patches and their resolution. Lower-resolution attention maps are restricted to fewer Fourier modes at high $k$ as a direct consequence of their reduced Nyquist frequency, which limits the range of scales over which correlations can be captured at such resolutions. This causes the model to focus on primarily the largest coherent structures in the density field. This is observed in Fig.~\ref{fig:lower_res_cross_power}, where the $16\times16$ and $8\times8$ layers exhibit significant correlations over a range shifted towards larger scales, spanning $\sim31$ - $125~h^{-1}\mathrm{Mpc}$ ($\sim0.05$ - $0.2~h\mathrm{Mpc}^{-1}$) and $\sim44$ - $125~h^{-1}\mathrm{Mpc}$ ($\sim0.05$ - $0.14~h\mathrm{Mpc}^{-1}$), respectively. This shift is consistent with the reduced Nyquist frequencies of the downsampled density fields, corresponding to $~15.71~h^{-1}\mathrm{Mpc}$ for the $16\times16$ case and $~31.42~h^{-1}\mathrm{Mpc}$ for the $8\times8$ case, reflecting the range of scales accessible to the model at each resolution. However, the scales at which the cross-power demonstrates strong correlations are also a consequence of the timestep at which the attention maps are generated in the trained diffusion model, which can be directly tied to the cutoff scale $\Lambda_{t}$, above which Fourier modes are smoothly suppressed. This is discussed in further detail in Appendix \ref{app_diffusion_timestep}.

    Consistent with the Dice analysis, the cross-power reveals that different attention layers exhibit strong correlations with distinct cosmic web environments in the matter density field. Layer $1$ displays a strong response predominantly to overdense structures, exhibiting positive correlations for filament- and particularly node-associated maps, while void maps show negative correlations. In contrast, layer $2$ captures negative correlations for all environment cases, indicating a strong sensitivity to underdense structures. Layer $21$ shows sensitivities to both overdense and underdense structures, although with weaker amplitudes relative to layer $1$. Finally, the aggregated attention displays strong correlations for node maps and strong anti-correlations for void maps at large scales. Overall, these results further reinforce the fact that attention maps from different layers and resolutions are sensitive to distinct, spatially coherent structures of the cosmic web, particularly at intermediate to large spatial scales.

    \paragraph*{Cross-power with cosmic web environment density fields.}
    We extended this analysis by computing the cross-power between environment-associated attention maps and environment density fields $P_{\delta_{\alpha}, A_{\beta}}$, rather than the matter density field $\delta_{m}$. This allows us to assess how each environment map is sensitive to individual cosmic web environments. Figure~$\ref{fig:cross_comp_cross_ps_layer_2_same_comb}$ shows the resulting cross-correlations for attention layer $2$, particularly between matching environment pairs. It is evident that void maps are strongly correlated with void regions, whereas filament and node maps are anti-correlated with overdense structures. This is consistent with our Dice and previous cross-power results (Fig.~$\ref{fig:dice_coefficients}$ and Fig.~$\ref{fig:cross_power}$), confirming that this layer is predominantly sensitive to underdense environments. All possible cross-environment combinations between environment density fields and attention maps for attention layer $1$ are displayed in Fig.~$\ref{fig:cross_comp_cross_ps_layer_1}$, which also align with our results in Fig.~$\ref{fig:cross_power}$.

    \subsubsection{Cross-environment attention analysis}
    
\begin{figure}[htbp]
        \centering
        \includegraphics[width=\columnwidth]{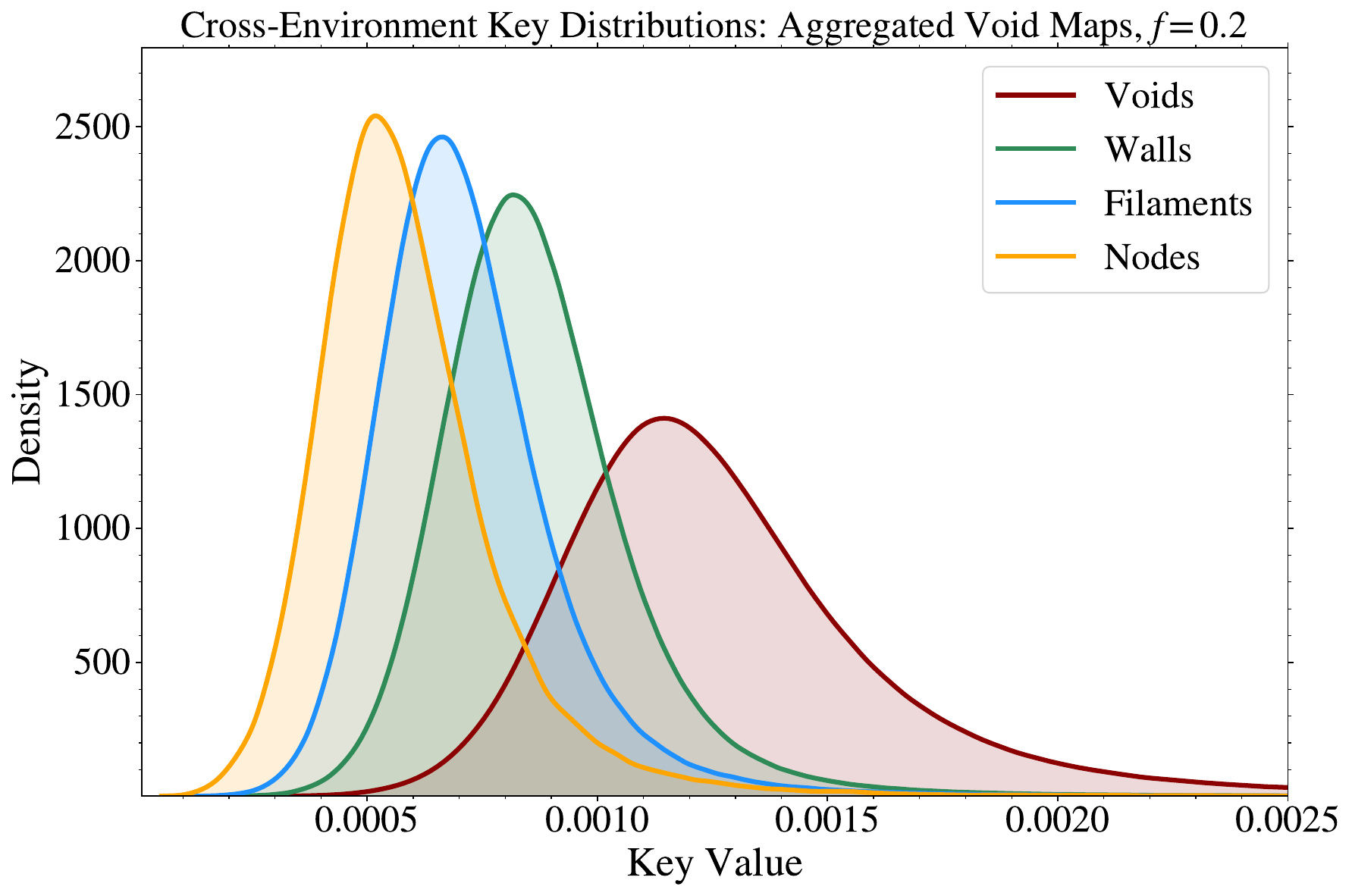}
        \caption{Environment-grouped key distributions of void-associated attention maps in the aggregated attention layer.}
        \label{fig:cross_comp_dist}
    \end{figure}

    Finally, we performed an analysis based on key distributions of the attention maps, as discussed in Sect.~\ref{sect_cross_comp}, to assess how strongly environment-associated maps respond to different environments, that is, how much void maps attend to void or filament regions, by comparing their key distributions. Given that the aggregated attention incorporates structural information across multiple layers and resolutions, providing a multi-scale representation of the cosmic web, we focused on this layer for our analysis. The resulting environment-grouped key distributions of $32\times32$ void attention maps are shown in Fig.~\ref{fig:cross_comp_dist}. 
    
    Void maps show a clear separation of the different environments, appearing to have a Gaussian-like distribution, with the highest key activation values belonging to void regions, followed by wall, filament and node regions. This behaviour indicates intra-attention similarity, where maps associated with a given environment respond most strongly to that same environment. Additionally, void maps assign the second highest activations to walls after voids, indicating a sensitivity to structurally related environments, as walls form the boundaries between them.
    
    The cross-environment attention analysis reveals that self-attention maps can capture not only specific cosmic web environments, but also the structural relationships between them. Through intra-attention similarity, attention maps give the strongest responses to environments matching their associated query environment, while still exhibiting a sensitivity to other environments.

\section{Discussion}\label{sec_discussion}
    Our statistical estimators consistently indicate that attention maps encode spatially coherent, multi-scale representations of cosmic web environments, with a dominant sensitivity to large-scale structures. The cross-power analysis shows that attention maps predominantly capture large-scale structural information (see Sect.~\ref{sect_cross_power_analysis}), with dominant positive and negative correlations between the DM density field and node and void maps, respectively. The Dice coefficient analysis further supports this, showing dominant negative correlations between void maps and the DM density field across multiple attention layers. This behaviour is consistent with \cite{lahiry2025interpretingcosmologicalinformationneural}, which showed that CNNs trained on hydrodynamical simulations obtain most of their constraining power from information extracted from large physical scales relative to the spatial extent of the input density fields. Specifically, their networks extract cosmological information from the most extreme overdense and underdense environments, with sensitivity primarily towards void-dominated regions as it constitutes most of the cosmic web. Similarly, \cite{Ullmo_2021} demonstrated that GANs and autoencoders trained on both native $2$D N-body simulations and separately $2$D slices of $3$D simulations encode large-scale structures effectively, while small-scale features are reconstructed less robustly. 
    
    \cite{aymerich2025interpretabilitydeeplearningmethodsapplied} found that CNNs trained on weak lensing maps extract the most information from structures at scales near the transition between linear and non-linear regimes at the considered redshifts, exploiting both Gaussian and non-Gaussian information. While these scales are small compared to the full extent of the simulation volumes ($500~\mathrm{Mpc}^{3}$) from which the weak lensing maps are created, their analysis highlights the importance of the multi-scale representation of cosmic web structures encoded by attention maps. Such an encoding helps the diffusion model capture both linear and non-linear features in the generated density fields.

    However, the specific physical scales at which significant correlations are captured by attention maps depend on the spatial extent and resolution of the input density field. Our analysis is restricted to a relatively narrow range of physical scales, approximately $7.85 - 125~h^{-1}\mathrm{Mpc}$. Therefore, input samples that span larger physical scales would enable the model to capture correlations at correspondingly larger scales, while higher-resolution input fields could improve the sensitivity of the attention maps, and thus the diffusion model, to smaller physical scales. This is evident in the cross-power analysis of lower resolution maps shown in Fig.~\ref{fig:lower_res_cross_power}, as discussed in Appendix~\ref{sect_preprocessing_cross_power}. This scale dependence is also consistent with the analysis discussed in \cite{lahiry2025interpretingcosmologicalinformationneural}, where the impact of removing large-scale modes depends on the simulation box size.
    
    The segmentation using T-Web is performed on full $3$D volumes to obtain the ground truth labels for cosmic web environments, whereas the diffusion model is trained on $2$D slices and learns semantic information present in the input $2$D plane only. This produces projection effects, where some query positions may appear to belong to a wall or a filament region in the projected field, despite the corresponding grid cell actually being classified under a void environment by T-Web. As a result, some attention maps fail to capture meaningful structural information and are instead dominated by noise. Furthermore, the limited spatial resolution of $32\times32$ patches can produce many attention maps, particularly those from lower-resolution layers, that exhibit significant noise. To mitigate noise in our analyses, we applied a threshold to the density fields, retaining the most characteristic regions of each environment (with $f=0.2$ corresponding to the lowest $20\%$ of pixel intensities for voids, the highest $20\%$ for nodes, and the middle $20\%$ for walls and filaments). This isolates the deepest voids, the densest nodes, and the mid-range density structures of walls and filaments. The attention maps of the retained pixels were then used for quantitative analysis, comparing against the overall DM density field. This threshold value was chosen as it provides the best balance between the reduction of projection effects and enhancing the signal of statistical estimators considered in our analysis. We verified in Fig.~\ref{fig:cross_ps_ps_agg_attn_multiple_alphas_fracs} (right panel) that the separation between environments, particularly in the cross-power amplitudes of the aggregated attention, persists across a range of threshold values ($f=0.1 - 0.4$), despite a decrease in overall magnitude of the amplitudes with increasing $f$. The $\pm3\mathrm{SEM}$ bands also remain clearly separated between environments for each case, confirming that the separation is not simply due to the strong selection of attention maps corresponding to retained pixels in extreme characteristic regions of environments. For the cross-environment attention analysis, we additionally applied the same threshold to the key positions, removing keys corresponding to spatial locations that are filtered out in the input density field. 
    
    By analysing attention maps with statistical estimators against T-Web segmented cosmic web environments, we find that the model captures non-Gaussian structural information, going beyond two-point statistics. This analysis can be naturally extended by incorporating additional summary statistics such as Minkowski functionals, which would allow for a more comprehensive evaluation of both attention maps and generated density fields, providing a deeper insight into the model's ability to capture non-Gaussian information. Moreover, applying this framework to a diffusion model trained on three-dimensional density fields would provide access to the full connectivity of cosmic web structures, providing a more accurate representation of filaments and walls and potentially reducing projection effects that contribute noise in the attention maps.

    The model checkpoint used for the self-attention analysis was chosen by evaluating the statistical convergence of the matter power spectrum between generated and true samples. We compared multiple checkpoints spanning the full training run of $10^{6}$ iterations (see Appendix \ref{model_arc_setup_section}) and adopted the checkpoint at $240{,}000$ iterations, which provides the best agreement. We carried out an additional independent training run with a different weight initialisation and carried out our cross-power analysis on the attention layers of the resulting model. We observe that the same qualitative behaviour is recovered across the different layers and the aggregated attention of this model. This confirms that, although the specific layers that capture particular environments may vary across independently trained models with the same configuration, the underlying physical structures captured by the self-attention mechanism remain consistent.

\section{Conclusion}\label{sec_conclusion}

    In recent years, generative diffusion models have shown that they are a promising tool for emulating cosmological simulations. However, the internal latent representations through which structures in the cosmic web are encoded remain largely unexplored. For this work, we presented a study combining qualitative and quantitative analyses of self-attention maps to assess their ability to capture cosmic web environments. We trained a diffusion model on two-dimensional slices of DM density fields from N-body simulations, from which we extracted attention maps across multiple layers and spatial resolutions to examine the semantic information that they learn. 
    
    To quantify how effectively attention maps capture cosmic web structures, we employed three statistical estimators: the Dice coefficient, cross-power spectra, and cross-environment attention distributions. The Dice coefficients demonstrate that node- and filament-associated maps overall respond primarily to overdense structures, while void-associated maps are sensitive to underdense regions. The cross-power analysis further reveals that the strongest correlations occur at intermediate-to-large scales relative to the spatial extent spanned by the input density field, with decreasing amplitudes towards smaller scales, indicating that attention maps primarily capture global, spatially coherent features of the cosmic web. Finally, the cross-environment key distributions show that the strongest responses occur in key positions belonging to environments that match their map's associated query positions (i.e. void regions have the strongest activations within void-associated maps). Clear separation of overdense and underdense structures is exhibited by the aggregated attention, which combines information from multiple layers of varying spatial resolutions.

    Overall, our analysis demonstrates that self-attention maps in diffusion models learn a multi-scale representation of cosmic web structures through the distribution of attention layers with various spatial resolutions throughout the model architecture. Lower-resolution maps capture global, large-scale structures due to their larger receptive fields, while higher-resolution maps capture more localised, small-scale structures. They demonstrate both intra-attention and inter-attention similarity, where key positions that share structural features with the map's associated query give strong responses, and maps whose queries belong to similar structures give similar activation responses. Generated density fields from our trained model reproduce two-point statistics with high accuracy, while visually displaying complex morphological cosmic web structures. This indicates that the model genuinely encodes non-Gaussian information, aided by the long-range dependencies captured by self-attention mechanisms in the model.

\begin{acknowledgements}
The authors thank the referee for their comments and suggestions, which helped clarify the paper. MN acknowledges financial support from the Centre national d’études spatiales (CNES), France (ROR: https://ror.org/04h1h0y33), as part of the PhD project "AI-Aided Cosmology", and from the Région Île-de-France through DIM Origines (IDF-DIM-ORIGINES-2024-1-01) under the "AI4Cosmo" project. The authors also thank the developers of the Quijote simulation suite for making their data publicly available.
\end{acknowledgements}

\bibliographystyle{aa}
\bibliography{myrefs}

\begin{appendix}
\section{Experiment}\label{setup_section}
    \subsection{Diffusion model architecture}\label{model_arc_setup_section}
    
    We utilised the diffusion model described in Sect.~\ref{RGDM_sec}, which consists of a U-Net architecture, comprising four downsampling and four upsampling blocks, connected by a bottleneck and skip connections linking the corresponding encoder and decoder blocks. Each downsampling block and the bottleneck contain two residual blocks, while each upsampling block includes three. Self-attention layers are embedded within each of these residual blocks, resulting in a total of $21$ attention layers: $8$ in the encoder, $1$ in the bottleneck and $12$ in the decoder. The spatial resolution of the maps in these layers vary across $r_{k}\in\{4, 8, 16, 32\}$. We also ensured that the convolutions in all CNNs throughout the U-Net operate under a `reflect' padding scheme, as this produces fewer noise artefacts in their attention map outputs, particularly compared to a padding of `zeros'.
    
    In our renormalisation-group interpolant, we adopted the same propagator $K_t(k)$ as in the 2D setting of \cite{masuki2025generativediffusionmodelinverse}, namely
    \begin{equation}
        K_t(k) = \frac1{\frac{k^2}{\Lambda_t^2}+1}~,
        \label{propagator_eqn}
    \end{equation}
    effectively suppressing smoothly the modes above the cutoff scale $\Lambda_t^2=\Lambda_0^2 e^{-2t/\tau}$. With our choice of noise schedule, the forward process converges at $t=T$ to a Gaussian field with covariance
    \begin{equation}
        G_0(k)=\frac{1}{m^2+k^2}~,
    \end{equation}
    where we set $m=0.15$ to roughly match the shape of the input spectra. In practice, the precise value of $m$ had little impact on the performance of the trained models.
    
    The number of diffusion steps $T$ determines the discretisation of both the forward and backward stochastic processes. Choosing $T$ too small degrades the quality of the generated samples, whereas larger values increase the computation cost of the generation. Based on the analysis of \cite{masuki2025generativediffusionmodelinverse}, we set $T=300$, as quality metrics were observed to plateau beyond this value for $32\times 32$ images. Considering a batch size of $256$, the model was trained over $10^6$ stochastic gradient descent updates, with checkpoints saved periodically. We implemented the weighted $L_{2}$ loss function for minimisation during training, given by Eq.~(\ref{rgdm_loss}). We evaluated the statistical convergence of the matter power spectrum between true and generated samples across multiple checkpoints throughout the training run, from which we find the best agreement to be at $240{,}000$ iterations. Hence, we adopted this model for our analysis of the self-attention.

    \subsection{Attention map extraction}\label{sect_attention_map_extraction}
    
    For our analysis, we considered $11$ attention layers in our trained model that visually capture prominent structural features: five high-resolution layers of dimensions $r_{{k} \in \{1,\ 2,\ 19,\ 20,\ 21\}}=32^{4}$, one low-resolution layer of dimensions $r_{5} = 8^{4}$, and the remaining five are of dimensions $r_{{k}\in\{3,\ 4,\ 16,\ 17,\ 18\}} = 16^{4}$. Layers excluded from these sets were motivated by several reasons: very low-resolution maps of several layers display higher noise levels compared to higher-resolution maps, due to their larger receptive fields that compress many pixels down into a single value. Due to the limited resolution of the input $2$D density patch itself, this effect further enhances the noise present at lower-resolution layers found deeper in the U-Net. Additionally, maps in layers from the decoder blocks also display more noise relative to those from the encoder blocks, as many noisy artefacts are introduced due to upsampling lower-resolution maps back to a higher-resolution. Thus, these excluded layers fail to effectively capture coherent structures.
    
    To construct the aggregated attention tensor, $9$ out of the aforementioned $11$ layers were utilised, as layers $19$ and $20$ were further excluded due to the presence of significant noise artefacts, improving the coherence of captured structures in the aggregated maps. The subset of $9$ attention layers $\mathcal{A}_{k}$ used for aggregation are given by $\mathcal{K}=\{1, 2, 3, 4, 5, 16, 17, 18, 21\}$.

    \subsection{Diffusion timestep}\label{app_diffusion_timestep}
\begin{figure}[htbp]
        \centering
        \includegraphics[width=\columnwidth]{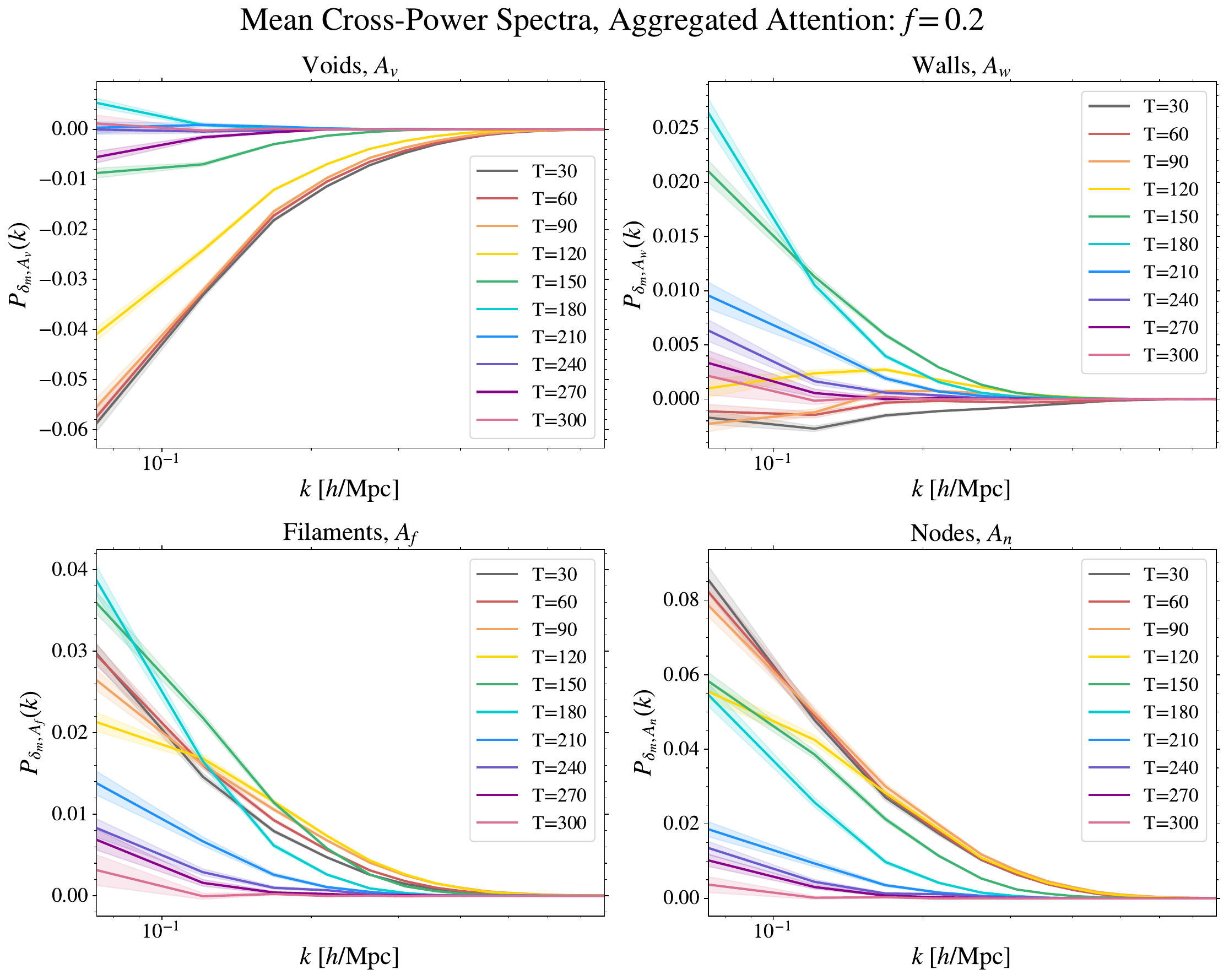}
        \caption{Cross-power spectra of the aggregated attention between environment-associated attention maps of thresholded queries ($f=0.2$) and the matter density field for varying diffusion timesteps. Shaded bands indicate $\pm\mathrm{SEM}$ of the mean cross-power over $500$ density field samples. A timestep of $T=60$ exhibits the most balanced cross-power amplitudes across cosmic web environments.}
        \label{fig:cross_ps_across_ts}
    \end{figure}

    To extract attention maps from all layers of the trained model, we provided it with an input log density field from the \textsc{Quijote} suite noised at a diffusion timestep of $t=60$, initially chosen based on visual inspection, as the resulting attention maps at this timestep best capture distinct and well-resolved cosmic web structures. This was further confirmed through a cross-power analysis computed across varying diffusion timesteps for the aggregated attention, shown in Fig.~{\ref{fig:cross_ps_across_ts}}, demonstrating the most balanced cross-power amplitudes for cosmic web environments at this timestep. Since the aggregated attention incorporates both local and global structural information from multiple layers across varying spatial resolutions, it was an appropriate choice for this analysis. The sensitivity of the self-attention to intermediate-to-large physical scales can be linked to the cutoff scale $\Lambda_{t}$ set by this timestep choice. Over the forward process, $\Lambda_{t}$ decreases monotonically, causing progressively larger physical scales (lower $k)$ to cross into the suppressed regime as $t$ increases. Correspondingly, Fig.~\ref{fig:cross_ps_across_ts} shows the scales at at which the cross-power exhibits strong correlations shifting towards lower $k$ as $t$ increases. At $t=60$, the cutoff scale corresponds to $\Lambda_{60}\simeq3.80\,h/\mathrm{Mpc}$, which lies beyond the Nyquist frequency $k_{\mathrm{Nyq}}$ of our $32\times32$ DM density fields. However, since the transition of preserved to suppressed scales defined by $K_{t}(k)$ is smooth, modes below $k_{\mathrm{Nyq}}$ also face some mild suppression, with $\sim95.71\%$ of the original signal being retained at $k_{\mathrm{Nyq}}$. Consequently, attention maps at this timestep focus on structures at intermediate-to-large scales, with gradual decay of the signal towards smaller scales, consistent with our cross-power analysis.
    
\section{Quantitative analysis}

    \subsection{Preprocessing of cross-power analysis}\label{sect_preprocessing_cross_power}
    For each cosmic web environment, we computed the cross-power spectra between $2000$ input density field samples and their associated attention maps across multiple layers with resolutions $r\in\{8, 16, 32\}$. To analyse lower-resolution maps, the density fields and the corresponding T-Web segmentation masks were downsampled (via max-pooling for the latter case) to match the attention map resolution. The downsampled masks were then used to provide environment labels for the queries associated to lower-resolution maps. Although the downsampled fields retain the same large scale modes as the original resolution, the Nyquist frequency is reduced (e.g. downsampling a density field from $32^{2}$ to $16^{2}$ reduces the Nyquist frequency $k_{\mathrm{Nyq}}$ from $0.80~h\mathrm{Mpc}^{-1}$ to $0.40~h\mathrm{Mpc}^{-1}$). To prevent aliasing effects, a Fourier-space low-pass filter was applied prior to downsampling to remove modes above the new Nyquist frequency.

    \begin{figure}[htbp]
        \centering
        \includegraphics[width=0.495\columnwidth]{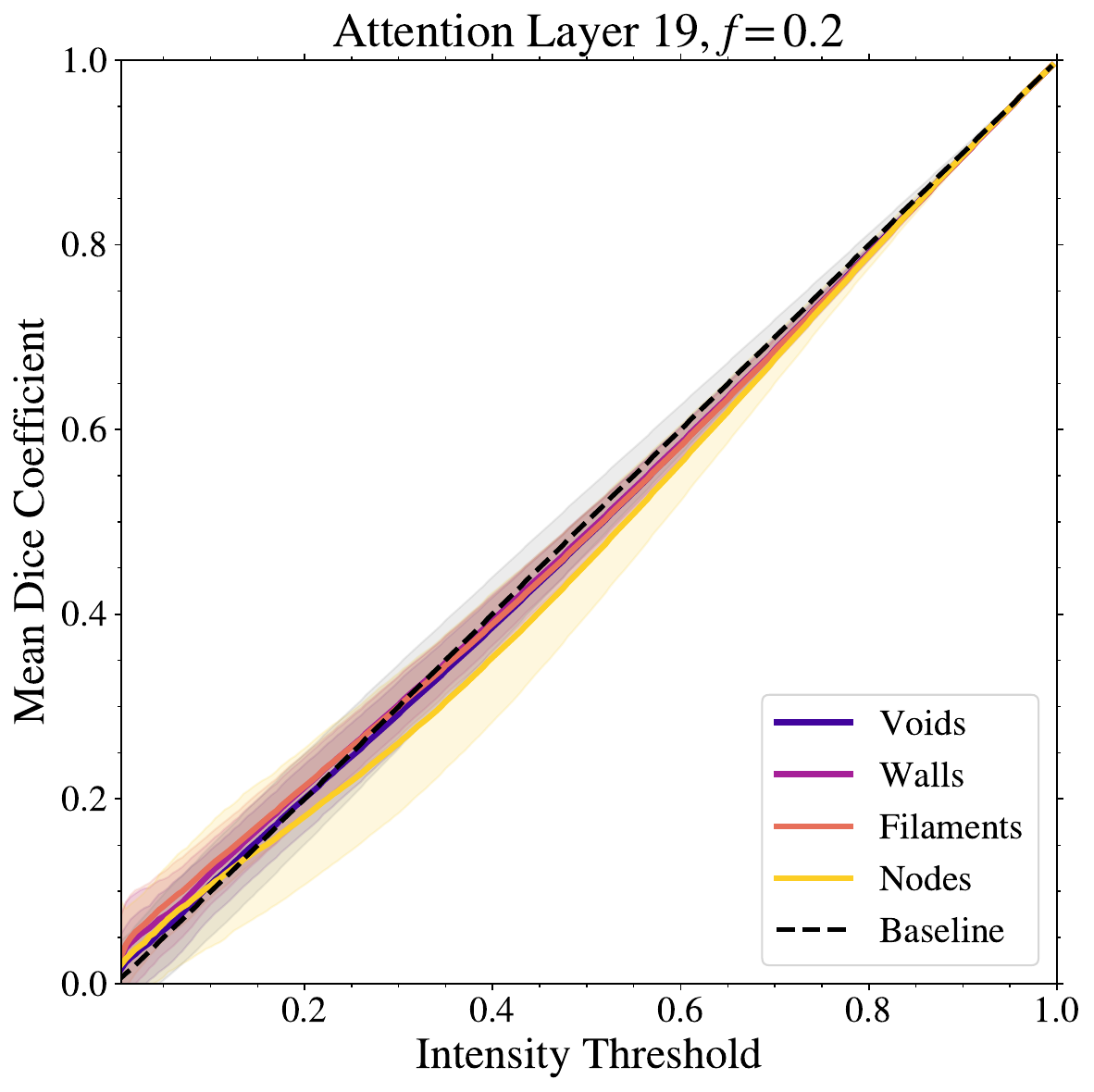}
        \includegraphics[width=0.495\columnwidth]{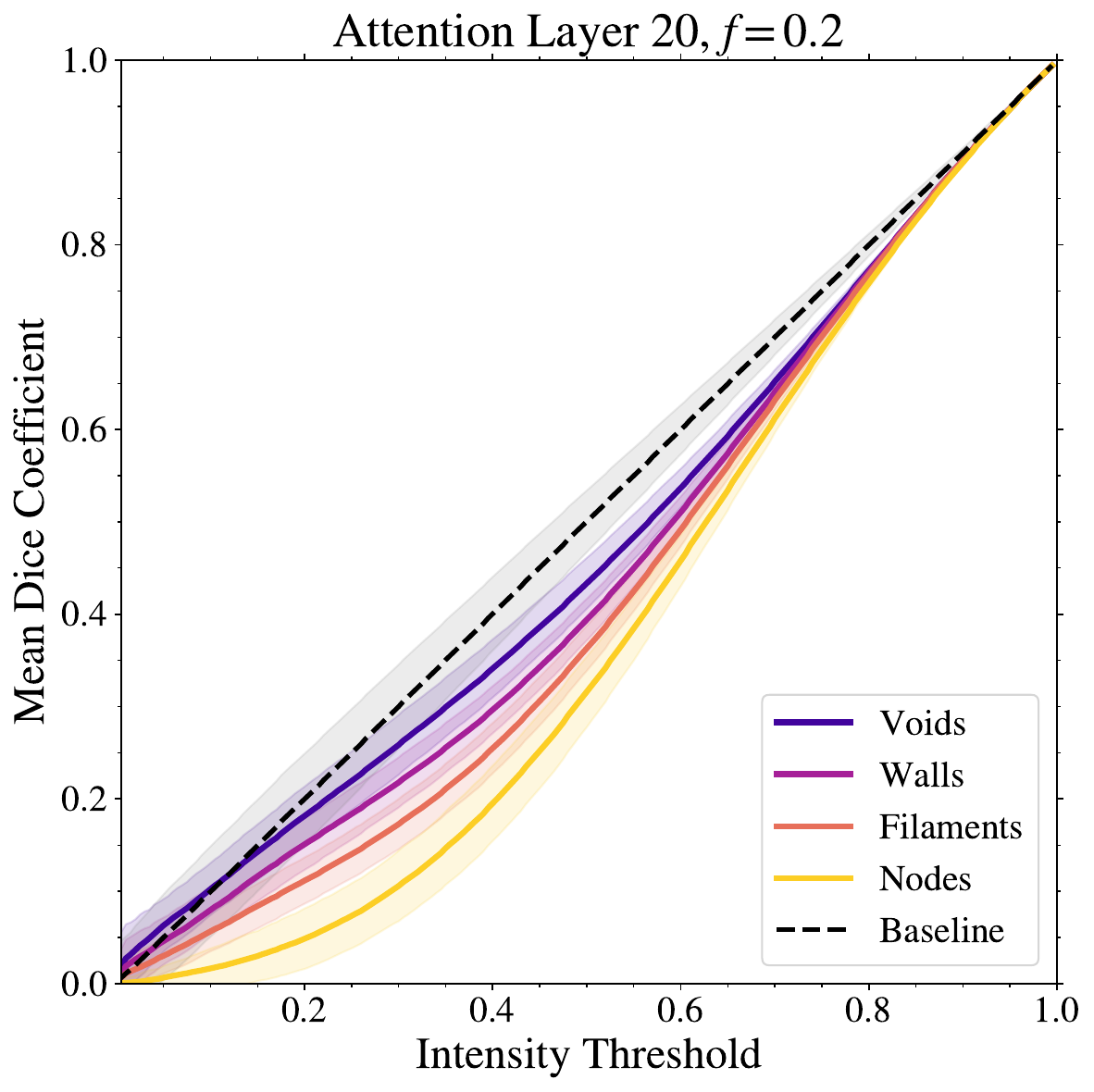}
        \caption{Dice coefficient curves for attention layers $19$ and $20$, quantifying the extent of structural overlap between attention maps of different cosmic web environments and their associated matter density fields across pixel intensity thresholds. Shaded bands represent the $1\sigma$ dispersion of the mean Dice coefficient across $200$ samples at each threshold. }
        \label{fig:dice_19_20}
    \end{figure}

    \begin{figure}[htbp]
        \centering
        \includegraphics[width=\columnwidth]{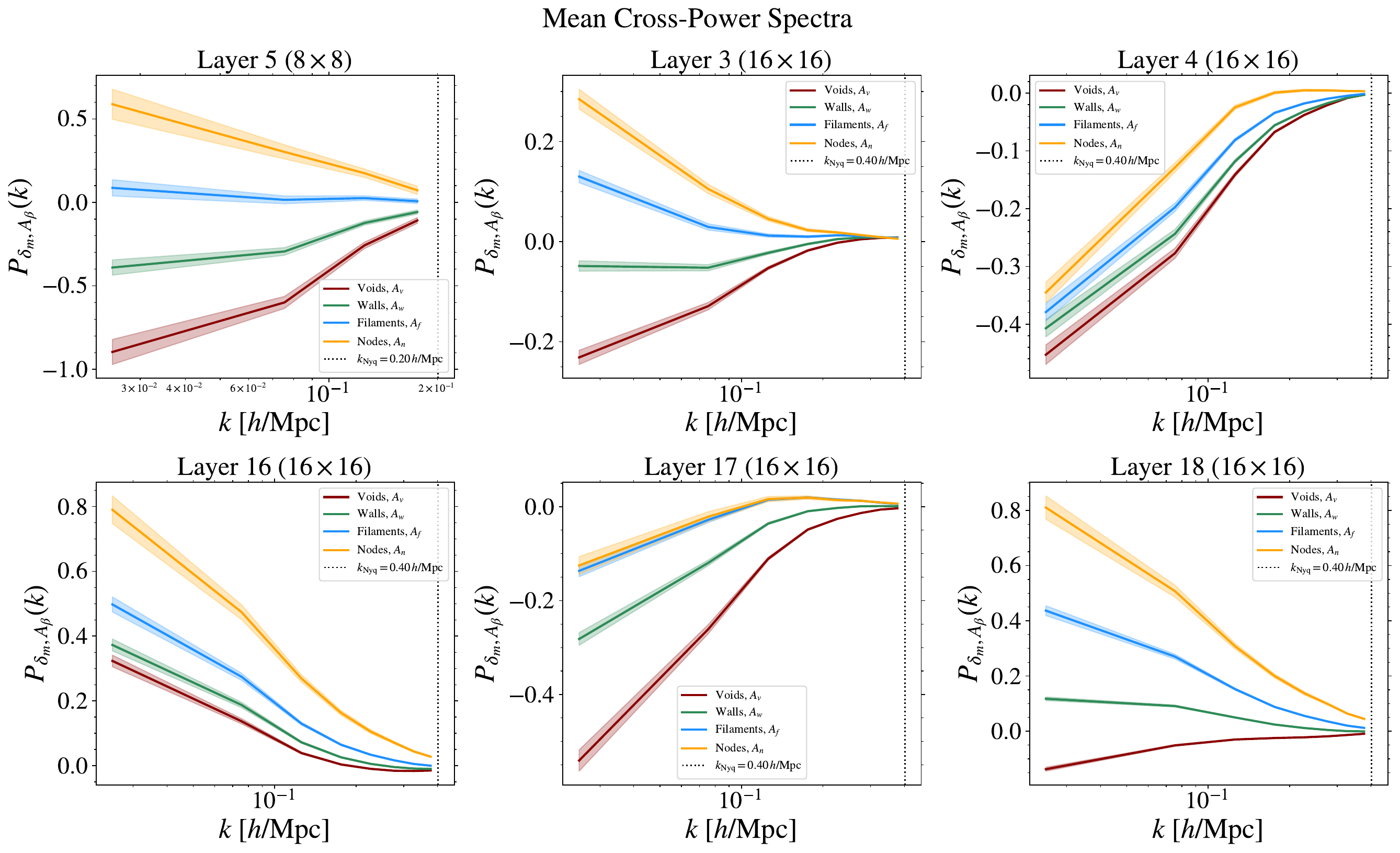}
        \caption{Cross-power spectra between downsampled log-transformed DM density fields and their matching resolution attention maps associated to different environments. No thresholding is applied to the maps. Shaded bands show $\pm3\mathrm{SEM}$ of the mean cross-power over $2000$ density field samples. Dotted vertical lines indicate the Nyquist frequency. We observe strong sensitivities to both overdense and underdense structures in layers $3$ and $5$, whereas layers $4$ and $17$ show strong attention responses to underdense structures overall for all environment cases, and layer $16$ showing an overall sensitivity to overdense structures. Layer $18$ captures strong correlations to overdense structures, specifically for the filament and wall environments, while also capturing anti-correlations with underdense structures for voids.}
        \label{fig:lower_res_cross_power}
    \end{figure}

    \begin{figure}[htbp]
        \centering
        \includegraphics[width=\columnwidth]{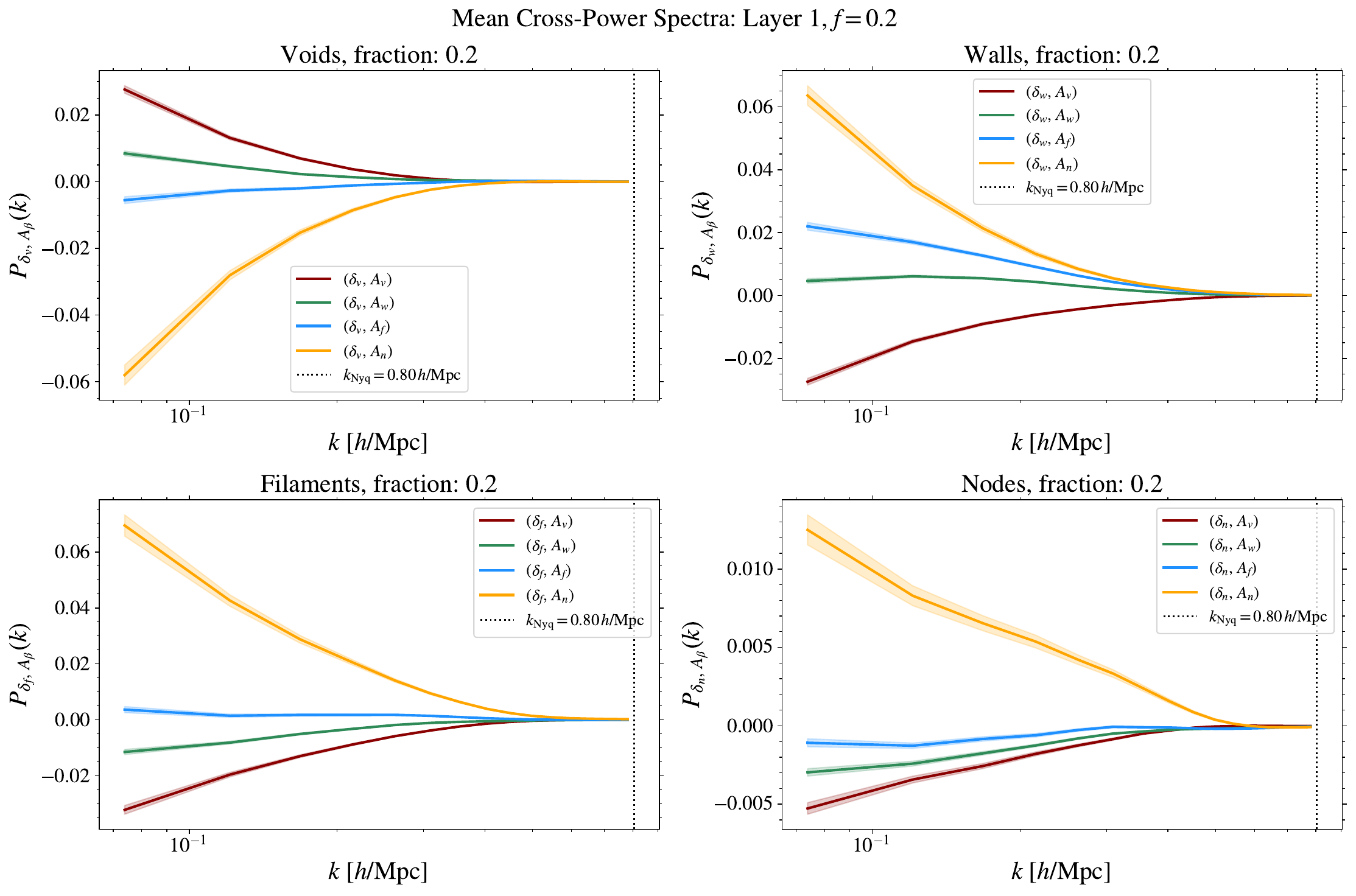}
        \caption{Cross-power spectra between all combinations of environment-associated attention maps and environment density fields. Shaded bands show $\pm3\mathrm{SEM}$ of the mean cross-power spectra over $2000$ density field samples. Dotted vertical lines indicate the Nyquist frequency. Maps associated with overdense environments exhibit positive correlations with overdense density fields and anti-correlations with underdense density fields, and vice versa.}
        \label{fig:cross_comp_cross_ps_layer_1}
    \end{figure}

    \begin{figure}[htbp]
        \centering
        \includegraphics[width=0.495\columnwidth]{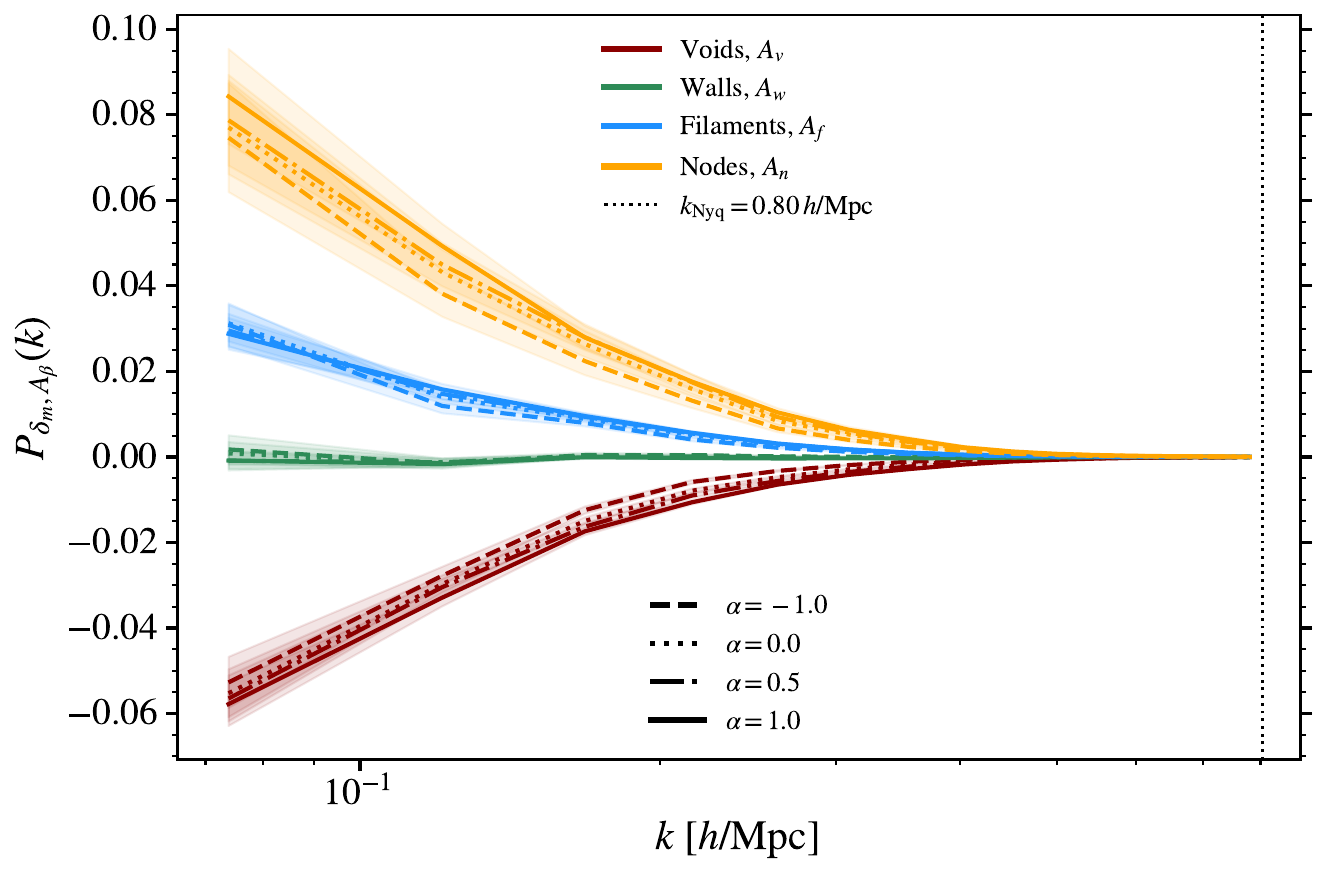}
        \includegraphics[width=0.495\columnwidth]{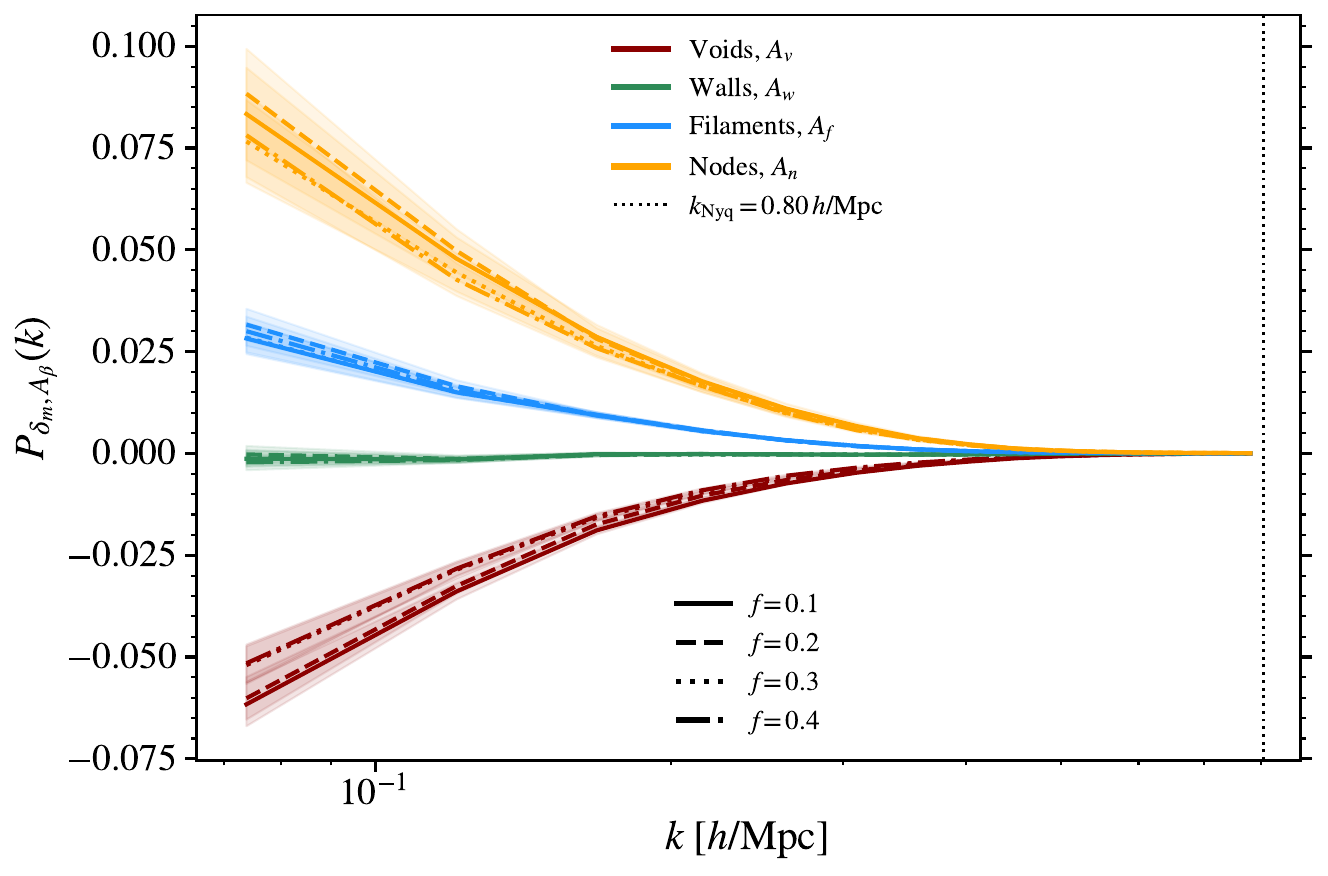}
        \caption{Cross-power spectra between log-transformed DM density fields and their corresponding $32\times32$ environment-associated attention maps from the aggregated attention layer. The shaded bands indicate $\pm3\mathrm{SEM}$ of the mean cross-power spectra over $500$ density field samples. The dotted vertical line indicates the Nyquist frequency. Left: Cross-power computed across multiple aggregation weighting exponents $\alpha$, at a fixed threshold of $f=0.2$. For $\alpha>0$, a stronger weighting to higher-resolution layers is introduced, whereas $\alpha<0$ gives a stronger weighting to lower-resolution layers. The overall magnitude of the amplitudes decreases progressively as $\alpha$ is reduced from $\alpha=1$ to $\alpha=-1$, particularly for node and filament maps, supporting the fiducial choice of $\alpha=1$ adopted throughout the main analysis. Right: Cross-power computed across multiple density thresholds $f$, for $\alpha=1$. Although there is a decrease in magnitude of the amplitudes with increasing $f$, the cross-power across different environments remains clearly separated for all threshold values.}
        \label{fig:cross_ps_ps_agg_attn_multiple_alphas_fracs}
    \end{figure}

    \begin{figure}[htbp]
        \centering
        \includegraphics[width=0.567\columnwidth]{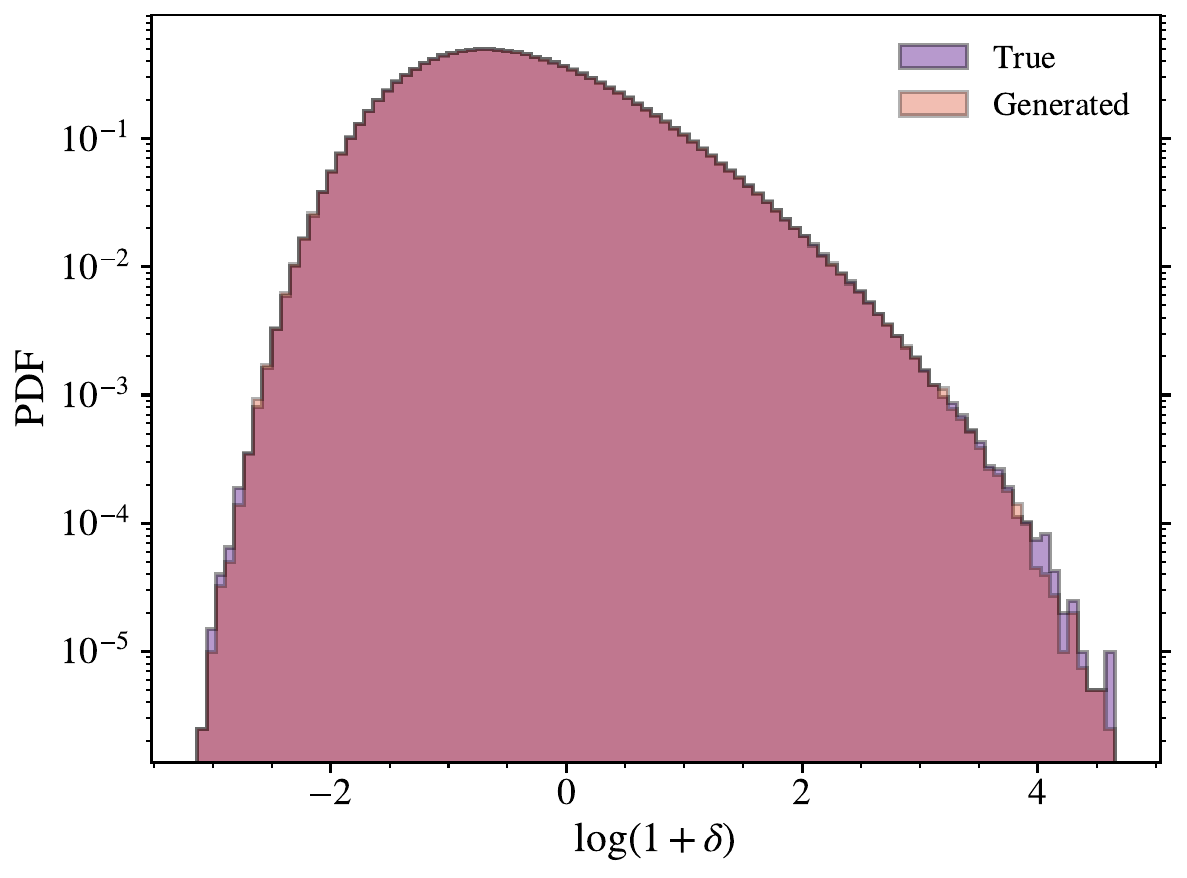}
        \includegraphics[width=0.424\columnwidth]{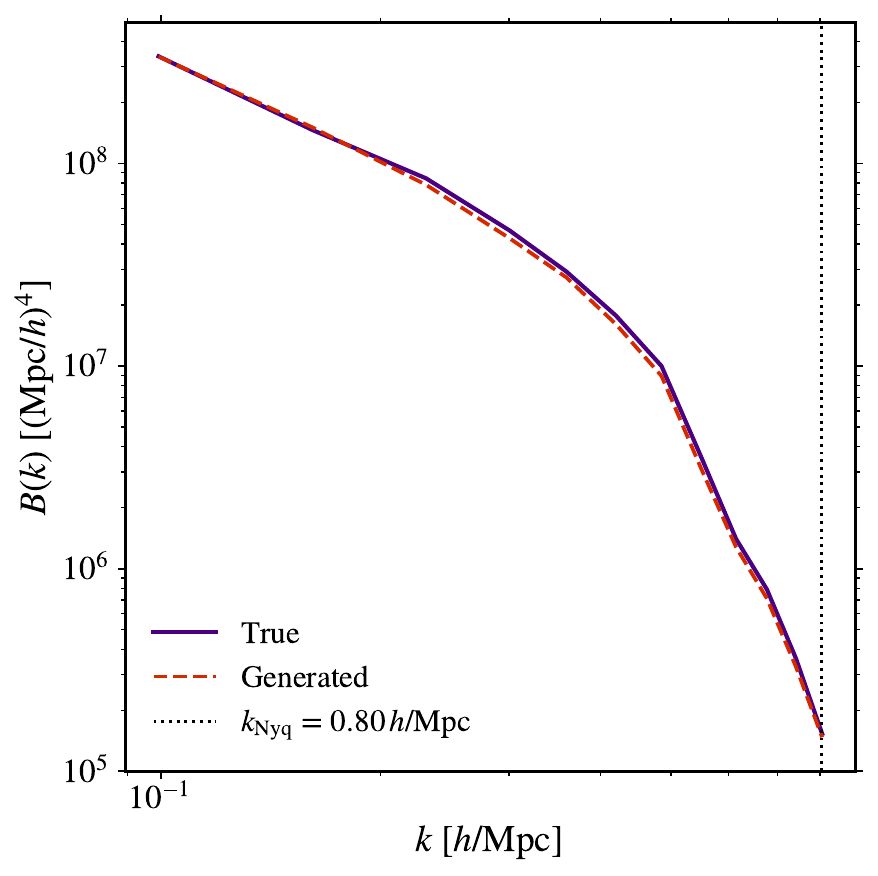}
        \caption{Comparison of higher-order statistics between true and generated DM density fields. Left: One-point PDF of $5{,}000$ true and generated $32\times32$ DM density field samples. The two distributions show an excellent agreement across the full range of pixel values, while exhibiting a clear departure from Gaussianity. Right: Mean equilateral bispectrum computed on $5{,}000$ density fields of both $32\times32$ generated (red) and true (purple) samples, whose amplitudes are closely in agreement with each other. The dotted vertical line represents the Nyquist frequency.}
        \label{fig:one_point_pdf_bispectra}
    \end{figure}

\end{appendix}

\end{document}